# A VISION FOR NUCLEAR THEORY

Report
of the
NSAC Subcommittee
on
Nuclear Theory

**11 November 2003**

Committee Members:

Joseph Carlson (LANL)
Richard Casten (Yale – ex officio)
Barry Holstein (U. Massachusetts)
Xiangdong Ji (U. Maryland)
Gail McLaughlin (NCSU)
Berndt Mueller (Duke – chair)
Witold Nazarewicz (U. Tennessee/ORNL)
Krishna Rajagopal (MIT)
Winston Roberts (Old Dominion/Jlab)
Xin-Nian Wang (LBNL)

# TABLE OF CONTENTS

# EXECUTIVE SUMMARY

This report responds to a request by the funding agencies (DOE and NSF) to the Nuclear Science Advisory Committee for review and evaluation of the currently supported efforts in nuclear theory and identification of strategies that ensure a strong U.S. nuclear theory program under various funding scenarios. In the 15 years since the last such report, often called the Koonin Report, nuclear theory has expanded vastly in scope. In addition to the traditional areas of nuclear structure and reaction theory, it encompasses wide areas of quantum chromodynamics relating to the structure of hadrons and the properties of hot and superdense nuclear matter, aspects of physics beyond the Standard Model, as well as the astrophysics of collapsed stars and the origin of elements. In spite of this dramatic expansion of the research field, the size of the active research community supported by the agencies has remained roughly constant, mainly due to a sharp decrease in the size of the nuclear theory program at the NSF.

Within its budgetary constraints, the DOE has given very good stewardship to its nuclear theory program over the past 15 years. The creation of the national Institute for Nuclear Theory in Seattle has helped to invigorate research in many areas of nuclear theory, fostered interactions with theorists from neighboring research fields, and raised the visibility of nuclear theory across science. The DOE program has seen a 15% growth in supported personnel, including graduate students, since 1986. The Outstanding Junior Investigator award program, introduced in 2000, has provided recognition and added support for the best young nuclear theorists.

On the basis of surveys of the research community – from theorists as well as experimentalists who were asked to answer detailed questionnaires – this report presents a representative compilation of recent research achievements and a list of important challenges for nuclear theory research. These challenges are not only opportunities for future research, but also essential contributions to the success of the scientific program laid out in the 2002 NSAC Long Range Plan for Nuclear Science. Our data, gathered from the surveys, the agencies, as well as expert witnesses, show that the realization of these opportunities requires a substantial increase in the nuclear theory research community and significant enhancements of the computational infrastructure available to nuclear theorists.

Our report makes several recommendations for how this can be achieved. For the growth in personnel, and the recruitment of the talented scientists required for this growth, the most urgent measures are the introduction of a postdoctoral prize fellowship program of national scope, a graduate fellowship program, and the creation of centers that can seed new faculty positions at universities and staff positions at national laboratories. In the optimal scenario, up to ten topical centers and five centers of excellence would be established, which would address research problems of critical relevance to the national program and take advantage of the existing interdisciplinary connections to address important problems of physics. We recommend that these centers be established by competitive proposals and that, although scientific excellence should be the primary review criterion, overall program relevance and leverage of the funds provided by the agencies should also be considered. We believe that these initiatives are so important for the future of nuclear physics that we recommend their implementation at a reduced level even in a constant level of effort scenario, although we are well aware that this will cause significant stress in the existing base program.

A second important component of our recommendations is the investment in computing facilities for the national user community to take advantage of timely opportunities in lattice gauge theory, simulations of core collapse supernovae, and *ab initio* many-body calculations. On a 3-4 year timescale, these applications have hardware requirements of the order of 10 Teraflops, but a sustained investment program in evolving state-of-the-art hardware will be needed toward Petaflops scale facilities a decade from now. These investments are absolutely essential for the success of the JLab, RHIC, and RIA physics programs and must be made even in a constant level of effort scenario if the nation wants to reap the returns from its past, present, and future investment in these much more expensive experimental facilities.

We also make a number of recommendations of lesser budgetary impact. We propose that the grant support per PI of NSF grants be raised to the same level as the average DOE grant (corresponding to a 50% increase) in order to eliminate existing imbalance in the support by the two agencies, even under a constant level of effort scenario. We recommend that the grants made to young researchers in the OJI program be substantially increased and that young scientists in national laboratories be allowed to compete for these awards. We recommend that the DOE make wider use of its ability, occasionally exercised in the past, to provide bridge support through grants for new faculty and staff positions. We anticipate that a substantial fraction of these bridging arrangements will occur in the context of the establishment of new research centers. Finally, we make suggestions for sabbatical support, as well as theory-led REU programs and summer schools.

Responding to the charge, we consider the implementation of these recommendations in three growth scenarios (23%, 39%, and 55% budget growth) and under a constant level of effort scenario. We document in this report the crucial opportunities lost, both for theory and for the entire nuclear physics program, in a constant level of effort scenario. Growth in the nuclear theory program, as strongly recommended in the recent long range plan, is essential to interpreting the data from present experimental facilities and to laying the groundwork for the future nuclear physics program. The report presents a detailed benefit analysis of the recommendations and initiatives. We urge the agencies to begin implementation of our proposed initiatives even in a constant level of effort scenario, to address as many of the new opportunities as possible, to showcase their effectiveness, and to strengthen the program by motivating highly talented young scientists to seek a career in nuclear theory. However, only the growth scenario (Option III) will realize the full benefits urgently needed for the successful execution of the scientific program described in the 2002 NSAC Long Range Plan.

# 1. Introduction

The Subcommittee was charged to help NSAC respond to a request from the agencies to review and evaluate current NSF and DOE supported efforts in nuclear theory and identify strategic plans to ensure a strong U.S. nuclear theory program under various funding scenarios. The charge letter pointed out that "among the opportunities and priorities identified in the 2002 NSAC Long Range Plan is an enhanced effort in nuclear theory and a large-scale computing initiative," and requested further guidance of "how available resources might be targeted to ensure that the needed theoretical underpinnings are developed to realize the scientific opportunities identified by the community."

The charge letter specified that the NSAC report "should document your evaluation of the present national program in theoretical nuclear physics and its effectiveness in achieving results in the science areas highlighted in the recent 2002 Long Range Plan. It should identify the scientific needs and compelling opportunities for nuclear theory in the coming decade in the context of the present national nuclear theory effort, and what the priorities should be to meet these needs." The agencies requested advice on an optimum nuclear theory program under funding scenarios of i) a constant level effort at the FY04 Nuclear Physics Congressional Requests and ii) with the increases recommended in the recent NSAC long range plan. The Subcommittee was asked to submit an interim report by September 2003 and a final written report by November 2003.

The highest priority in the 2002 NSAC Long Range Plan for Nuclear Science (see http://www.sc.doe.gov/henp/np/nsac/docs/LRP_5547_FINAL.pdf for the full report) was given to the following **Recommendation 1**:

> *Recent investments by the United States in new and upgraded facilities have positioned the nation to continue its world leadership role in nuclear science. The highest priority of the nuclear science community is to exploit the extraordinary opportunities for scientific discoveries made possible by these investments. Increased funding for research and facility operations is essential to realize these opportunities.*
>
> *Specifically it is imperative to*
> - *Increase support for facility operations – especially our unique new facilities, RHIC, CEBAF, and NSCL – which will greatly enhance the impact of the nation's nuclear science program.*
> - *Increase investment in university research and infrastructure, which will both enhance scientific output and educate additional young scientists vital to meeting national needs.*
> - *Significantly increase funding for nuclear theory, which is essential for developing the full potential of the scientific program.*

The recommendation to substantially increase the funding for nuclear theory is not a new one. It echoes a core recommendation of the 1988 report to the NSAC by the Subcommittee on Nuclear Theory chaired by S.E. Koonin (the "Koonin Report"), which emphasized the need to increase funding for nuclear theory by 70% over 5 years, from 6% of the total agency support for nuclear physics to 10%. Other recommendations of the Koonin Report were:

- Add 60-65 additional PhD level personnel over 5 years.
- Encourage creation and support theory groups at universities with strong experimental programs.
- Create one or more nuclear theory centers of truly national and interdisciplinary character.
- Include theoretical funding as integral part of new large experimental projects.
- Ensure adequate access to supercomputers and workstations.
- Foster interactions between nuclear theorists.
- Add nuclear theorists as permanent program officers.

Some of these recommendations have been implemented – most conspicuously and successfully the creation of a national Institute for Nuclear Theory in Seattle – but the call for a substantial increase in the research community has not been realized. We will present a detailed status report, including an update on several tables from the Koonin Report, in Section 2. As these data show, funding for nuclear theory at the DOE has grown since 1988 at a rate above nominal inflation, but this growth has been offset by a significant decline of the NSF nuclear theory budget. Overall, the field has not experienced a substantial increase in the number of supported scientists.

The Subcommittee collected information about the status of the field mainly in two ways: (1) by requesting input from the U.S. nuclear theory community through questionnaires, and (2) by requesting program-related data from the funding agencies. A very detailed questionnaire (reprinted in full in Section 9.1) was sent to all leading PIs on active NSF and DOE grants in nuclear theory. The PIs were asked to distribute the questionnaire to all co-PIs on their grant. We received a total of 79 written responses. A summary analysis of the community response to the individual questions is given in Section 9.1. In addition, an abbreviated questionnaire was sent to the leading PIs on all active experimental grants. The text of this questionnaire and a summary analysis of the 22 written responses can be found in Section 9.2. We also sent a brief questionnaire to all members of the Division of Nuclear Physics of the American Physical Society, soliciting input from theorists who had not been reached in the mailing to PIs on active grants. Only 4 responses were received.

Program managers at NSF and DOE were very responsive to our requests for funding data. This required a considerable investment of time on their part, because many data we asked for were not kept in readily accessible databases or were not categorized exactly in the way we needed them and had to be gathered "by hand." We especially thank Sid Coon, Brad Keister, and Earle Lomon for the work they did in support of our task. The data obtained from the funding agencies made it possible to update the most important tables contained in the Koonin Report, which allowed us to track the develop-ment of nuclear theory funding from 1987 on.

These tables and other relevant data are analyzed and discussed in Section 2. A short selection of recent research achievements is presented in Section 3; a representative list of research opportunities for nuclear theory is discussed in Section 4. We note that these sections are heavily influenced by the feedback we received from the community in response to our questionnaires, but they also reflect – by necessity – our personal biases. None of these lists is intended to be complete. We apologize in advance for omissions of achievements or opportunities considered important by other members of the nuclear physics community.

Our list of recommendations and new initiatives is presented and justified in Section 5; their possible implementation under various funding scenarios – as requested in the charge to the Subcommittee – is discussed in Section 6. The potential benefits of our recommendations and initiatives are analyzed in Section 7. The Appendices (Sections 8 and 9) contain relevant material, such as the full text of the charge to NSAC, the text of the questionnaires, an analysis of the responses, and a brief summary of the Subcommittee activities.

# 2. Status of the Field

## 2.1. Intellectual status of the field

Nuclear theory addresses many of the big intellectual questions in physics. Both hadron structure and nuclear structure theorists wrestle with the challenge of understanding how complex quantum systems emerge from simpler building blocks and, complementarily, how patterns and regularities arise in complex systems. Nuclear astrophysicists aim to understand the origin of the elements, one of the "Eleven Science Questions for the New Century" identified in the National Research Council report "Connecting Quarks with the Cosmos" (the Turner Report). Another one of these challenges is also central to our field, namely the quest to understand the properties of matter at exceedingly high density and temperature. Nuclear theorists use QCD to provide an *ab initio* understanding of the properties of hot matter of the sort that filled the universe during its first microseconds and that experimentalists seek to recreate in heavy ion collisions. And, having understood the properties of arbitrarily dense matter from first principles, they are currently wrestling with how the densest matter in the universe -- that at the core of neutron stars -- can be described from this starting point. Nuclear theory plays an important role in addressing two more of the eleven challenges, via elucidating neutrino oscillation phenomena in the sun and in supernovae, and via analysis of the implications for the baryon asymmetry of the universe of increasingly stringent limits on electric dipole moments of the neutron, nuclei, and atoms. Looking beyond today's challenging questions, our rapidly developing and hard-won understanding of the QCD vacuum, its hadronic excitations, and its phase transitions will likely be a prerequisite to understanding the dynamics and the cosmology of whatever physics lies beyond the Standard Model, assuming that it has new strong interactions at some very high energy scale.

Over the past 15 years, nuclear physics worldwide, and especially in the United States, has gone through an unprecedented transformation. In terms of research facilities, the field has changed from an infrastructure that was dominated by small and mid-sized research facilities to one dominated by two very large facilities (CEBAF and RHIC) whose operation accounts for almost half of the total nuclear physics budget at DOE. The NSF program is much smaller and supports only one national facility, the NSCL. The physics programs of these facilities are carefully planned and scrutinized by committees, which demand that the expensive experiments address well formulated questions. The role of nuclear theorists in motivating and justifying new experiments, and in analyzing and interpreting the experimental results from these facilities is larger than ever before.

At the same time, the scope of nuclear theory has broadened tremendously. A large fraction, roughly one half, of the theory community is now engaged in the study of quantum chromodynamics (QCD), as opposed to only a small vanguard of nuclear theorists at the time of the Koonin Report. Nuclear theorists study problems ranging from the electric dipole moment of the neutron, to the structure and spectrum of hadrons, the microscopic structure of nuclei such as $^{12}$C or $^{132}$Sn, hot and dense matter created in collisions of Au nuclei, and supernova explosions. Over the past decade, the techniques for a rigorous microscopic treatment of the structure of nuclei, hadrons, and hot nuclear matter have been developed. There is a tremendous need in the physics program laid out in the 2002 Long Range Plan for such calculations, presenting a plethora of outstanding research opportunities. However, as the community responses to our questionnaire have made clear, the existing number of theorists is not sufficient to attack even the most important problems with the required force.

## 2.2. Funding and personnel data

Here we will summarize the information about agency budgets and trends of funding. We also present updates of selected data contained in the Koonin Report (KR).

**Table 1: Number of personnel directly supported by NSF or DOE funds**

|  | Faculty/Staff | | Postdocs | | Total Ph.D. | | Grad. Students | |
|---|---|---|---|---|---|---|---|---|
|  | FY86 | FY02 | FY86 | FY02 | FY86 | FY02 | FY86 | FY02 |
| NSF Univ. | 62 | 39 | 21.5 | 10 | 83.5 | 49 | 32 | 30 |
| DOE Univ. | 80 | 103 | 43 | 44 | 123 | 147 | 72 | 95 |
| Total Univ. | 142 | 142 | 64.5 | 54 | 206.5 | 196 | 104 | 125 |
| Nat. Lab. | 46 | 37 | 18 | 26 | 64 | 63 | 7 | 3 |
| Total | 188 | 179 | 82.5 | 80 | 270.5 | 259 | 111 | 128 |

Table 1 lists the number of personnel (senior, postdocs, and graduate students) that are currently supported by grants or contracts from the agencies. The DOE nuclear theory program presently supports 44 programs at universities, 7 programs at national laboratories, and the Institute for Nuclear Theory. Overall, there is little change compared to the manpower situation at the time of the Koonin Report, but significant shifts have occurred in the distribution between the two agencies and between university and national lab groups. The substantial increase in the number of DOE-supported faculty and students has been offset by a dramatic decline in the number of faculty and postdocs supported by the NSF. In the DOE program itself, a 25% growth in the number of supported university faculty is in stark contrast with the 20% decline in the number of staff positions at national laboratories.

Table 2 gives the distribution of grant support by DOE for nuclear theory research divided by subfield, using the five Long Range Plan categories. Budgets are separately listed for universities and national laboratories. Funding from the SciDAC initiative is shown separately. About 45% of the DOE nuclear physics budget supports research in national laboratories; 55% supports university-based research.

**Table 2: DOE Nuclear Theory Support by Subfield (FY03 in k$)**

|                      | Universities | Nat. Labs | SciDAC | Total  |
|----------------------|--------------|-----------|--------|--------|
| Hadron Structure     | 3,213        | 2,674     | 1,178  | 7,065  |
| Nuclear Structure    | 3,272        | 2,698     |        | 5,970  |
| Hot Nuclear Matter   | 2,816        | 2,565     |        | 5,381  |
| Nuclear Astrophysics | 952          | 442       | 802    | 2,196  |
| Beyond SM            | 443          | 137       |        | 580    |
|                      |              |           |        |        |
| Total                | 10,696       | 8,516     | 1,980  | 21,192 |

Table 3 (corresponding to KR Table 2) presents a comparison between FY86 and FY03 of the fraction of DOE supported effort in the five Long Range Plan categories. Because the terms used to define various areas within nuclear physics have changed in the time since the Koonin Report, in the FY86 column we had to combine several smaller categories used in the Koonin Report in order to make the comparison with FY03. As the table demonstrates, significant growth of the efforts devoted to nuclear astrophysics and hot nuclear matter is counterbalanced by strong declines in the efforts directed toward nuclear structure and physics beyond the Standard Model. Hadronic physics has seen moderate growth.

**Table 3: DOE effort by subfield (%)**

|  | FY86 | FY03 |
|---|---|---|
| Hadron structure | 27.5 | 33 |
| Nuclear structure | 46.5 | 28 |
| Hot nuclear matter | 15 | 25 |
| Nuclear astrophysics | 4 | 10 |
| Beyond SM | 6 | 3 |

Table 4 (corresponding to KR Table 6) shows the change in the support for nuclear theory by the agencies over the past 15 years. Discounting inflation (we have used the Office of Labor Statistics tables, ftp://ftp.bls.gov/pub/special.requests/cpi/cpiai.txt) the programmatic support of nuclear theory by the DOE both at universities and the national laboratories has increased by 30% since 1987. This is quite remarkable in view of the fact that the manpower tables do not reveal a concomitant increase in FTEs over the same period. Indeed, some of this increase in funding at the national laboratories is spurious, corresponding to the recent transfer of operating funds (totaling about $2M) that had been traditionally used to support theoretical research into the nuclear theory budget. The remaining discrepancy between funding data and personnel data indicates that over the 16-year period from FY87 to FY03 a "constant level of effort" has required a budgetary increase of about 1% annually above rate of inflation reflected in the official CPI tables.

**Table 4: Annual funding for nuclear theory (in M$)**

|  | FY87 (1987 $) | | | FY87 (2003 $) | | | FY03 (2003 $) | | |
|---|---|---|---|---|---|---|---|---|---|
|  | NSF | DOE | Total | NSF | DOE | Total | NSF | DOE | Total |
| Universities | 2.6 | 5.5 | 8.1 | 4.2 | 8.9 | 13.1 | 2.9 | 11.7 | 14.6 |
| National Labs |  | 4.5 | 4.5 |  | 7.3 | 7.3 |  | 9.5 | 9.5 |
| INT (Prog.) |  |  | 0 |  |  | 0 |  | 0.9 | 0.9 |
| Total | 2.6 | 10.0 | 12.6 | 4.2 | 16.1 | 20.3 | 2.9 | 22.1 | 25.0 |

It is noteworthy that only 1/6 of the $6M increase (after inflation) of the DOE nuclear theory budget supports the workshop and visitor program of the Institute for Nuclear Theory; 5/6 has gone to bolster the base program. (More than half of the total INT budget of $2.1M pays for personnel, including senior scientists, fellows, and postdocs.)

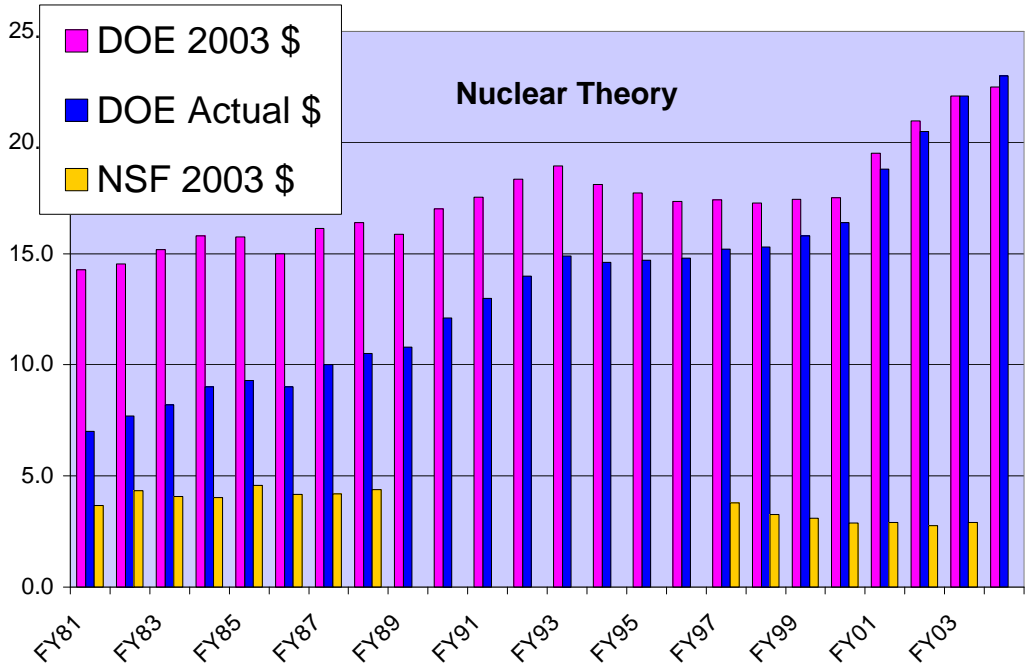

Figure 1: Nuclear theory budgets of NSF and DOE (1981-2004)

The 30% drop (after inflation) of the NSF nuclear theory budget is in stark contrast to the DOE numbers. In view of the discussion above, it is probably correct to assume that this corresponds to an effective drop of at least 40% in the level of effort within the NSF section of the program, as witnessed by the decline in personnel numbers.

The evolution of the nuclear theory budgets year by year is shown in Figure 1. The rapid increase in the DOE budget between FY00 and FY03 is, as already explained, largely due to the rebudgeting of funds from laboratory operations. Because the NSF did not track a specific budget category for nuclear theory before 1998, we were unable to fill in the gap between FY88 and FY98.

**Table 5: Grant support per senior Ph.D. (in k$)**

|  | FY86 (1986$) | FY86 (2003$) | FY03 (2003$) |
|---|---|---|---|
| NSF Univ. | 42 | 70 | 73 |
| DOE Univ. | 71 | 118 | 114 |

Table 5 (corresponding to KR Table 8) shows that the difference in support per PI or co-PI between NSF grants and DOE grants to universities has not been substantially reduced since the time of the Koonin Report. Per senior person, DOE grants are more than 50% larger than NSF grants. This imbalance has led to an increasing ineffectiveness of many NSF supported groups, but it has produced an even more deleterious effect: Talented young researchers submitting their first grant proposals as faculty members rarely consider the NSF as their funding agency of choice. In fact, we know of cases where a young theorist, faced with the option of accepting either a NSF or a DOE award, chose to reject the grant award from the NSF out of concern over the long-term outlook, which is reflected in Table 5.

## 2.3. Structural issues

### 2.3.1. Dynamics of the research field

The U.S. nuclear theory community has openly embraced the many new scientific challenges of the field and moved rapidly toward the frontiers. In the search for outstanding talent it has attracted many brilliant young theorists trained in high energy theory, in astrophysics, or abroad. Highly visible fellowship programs at the INT and the Riken-BNL Center have facilitated the entry of outstanding young investigators into the permanent ranks, and extensive programs providing bridge support have greatly helped in building strong theory communities interested in the physics explored at JLab and RHIC. The field is young: one-half of the responding PIs in our survey were less than 20 years past their Ph.D. degree, and this does not even include all those who are still in postdoctoral positions. The field is productive: the average number of students graduating each year with a degree in nuclear theory is about 35, eventually competing for about 5-7 faculty and staff positions annually.

**Table 6: Professional nuclear scientists at top-ranked physics departments***

| University | Theorists 1987 | Theorists 2003 | (Nearby) 2003 | Experimentalists 1987 | Experimentalists 2003 |
|---|---|---|---|---|---|
| UC Berkeley | 0 | 0 | 1 | 3 | 1 |
| Caltech | 1 | 2 | 1 | 5 | 3 |
| Chicago | 0 | 2 | 0 | 2 | 1 |
| Columbia | 0 | 1 | 4 | 3 | 3 |
| Cornell | 0 | 0 | 1 | 0 | 0 |
| Harvard | 0 | 0 | 1 | 2 | 1 |
| Illinois | 5 | 2 | 2 | 10 | 9 |
| MIT | 7 | 5 | 2 | 8 | 9 |
| Princeton | 0 | 0 | 0 | 8 | 2 |
| Stanford | 0 | 0 | 0 | 3 | 1 |
| Total | 13 | 12 | 12 | 44 | 30 |

*For purposes of comparison, we show the data for the same group of 10 universities as the Koonin Report. The 1993 NRC ranking includes UC Santa Barbara among the top-ten institutions. We note that the ranking of academic departments is not an exact science and that we do not consider this list as comprehensive.

The population of scholars in highly ranked physics departments is an important measure for the health and vitality of a field, because of the high visibility of these faculty positions and the strong concentration of the best graduate students in such departments. The Koonin Report had noted that the representation of nuclear theorists in the top-ranked departments had been in steady decline. Table 6 shows that this decline has modestly continued in the intervening years, although theorists appear to have fared better than experimentalists. It should be noted, however, that 3 of the 12 theorists listed in 2003 are on soft money positions, and two are astrophysicists whose research is supported in part by the nuclear theory program. For completeness, we have also listed the number of "nearby" theorists at these institutions, comprising theorists in other fields whose research connects to nuclear theory and some of whom have educated graduate students who have become nuclear theorists.

*2.3.2. Role of national laboratory groups*

Almost half of the DOE nuclear theory budget supports theory groups at national laboratories. Although the DOE support for laboratory theory groups has grown overall, several of the laboratory groups have suffered significant budget erosion over the past decade. This development has inhibited the overall physics efforts at national laboratories and has had a negative impact on the nuclear physics program nationwide. The national laboratory groups serve at least three distinct and important missions:

*Theoretical support for experimental programs*: Theory groups at national labs often provide direct support for the local experimental groups. Even at laboratories without a large facility, the experimental groups can be sizable and diverse. The effectiveness of these experimental groups can be strongly enhanced by a commensurate theoretical presence. The theory effort can be directed towards immediate experimental requirements, such as studies of radiative corrections to nucleon form factors measured at JLab. The scope of support is broader than this, however, as it can also be directed toward the development of new facilities such as RIA and NUSL.

*Addressing laboratory and national needs:* The national laboratories serve important and distinct purposes; within each laboratory specific needs often arise that require a strong theoretical presence. These needs can range from supporting other (non-nuclear) laboratory efforts to interfacing the nuclear theory community with national priorities. The laboratories cannot function effectively without a theory presence; in many cases the laboratories recognize this by providing institutional support on a short- or long-term basis to enhance the nuclear theory program.

*Special laboratory capabilities:* Each national laboratory offers a unique environment for research that may be difficult to sustain at universities. Laboratory programs can have a much larger scope than possible at most universities, allowing groups to focus a dedicated effort in specific areas or to cross disciplinary boundaries. In particular, laboratory environments are crucial for long-term projects that exceed the time scales of graduate education and tenure review, or for projects that require an exceptional degree of technical specialization. It must also be emphasized that such long-term projects are an essen-tial part of the portfolio of the DOE, running in parallel with many of the long-term experimental projects. Many of the achievements documented in this report have been carried out with significant contributions from national laboratory groups. For example, many achievements with a large computational component, including lattice QCD, nuclear astrophysics, and quantum many-body simulations have been attained with important contributions from researchers at national laboratories.

These arguments demonstrate why the theory programs at all national laboratories must be supported at levels that allow effective research to be carried out. We believe that the needs of the national laboratory groups can be addressed within a coherent set of theory initiatives. These initiatives are designed to foster effective collaborations between university and national laboratory groups to obtain the largest positive impact on the nuclear physics program.

*2.3.3. Institute for Nuclear Theory*

The National Institute for Nuclear Theory was created by DOE following the recommendation of the Koonin Report. The INT has been recognized widely, both inside and outside of our community, as a huge success. It creates a unique environment where researchers from different institutions and backgrounds can meet, working closely together on burning physics problems, and thereby fosters new ideas and collaborations.

Every year, the INT runs three roughly three-month long programs with typically 60-90 visitors each, as well as two shorter workshops. These programs, which are selected from the community's proposals by the INT's National Advisory Committee, have played a significant role in stimulating and advancing research in nuclear physics and have had great intellectual impact on the field as a whole. Many new frontier areas and experimental facilities have been the subject of INT programs responding to the needs in the field. There are a number of examples where INT programs have led to the development or blossoming of significant new research directions. The INT has helped to enhance the recognition of our field by fostering high quality interdisciplinary programs which overlap with astrophysics, atomic physics, condensed matter physics, and particle physics. The INT programs contribute to strengthening the international ties in our field, since 35-40% of the program visitors come from overseas.

The INT also runs joint workshop series with Jefferson Lab and RHIC, which address important and timely theoretical issues related to ongoing experiments at these facilities and promote interactions between theorists and experimenters. A new workshop series related to RIA physics has recently been launched. In addition, the INT sponsors various workshops at other institutions.

An important contribution of the INT is in the nurturing of young theorists. The broad intellectual scope of the INT provides an ideal environment for young people to further their scientific development and helps them get the recognition they deserve. The success ratio has been high: a large fraction of the postdocs go on to faculty or staff positions.

The INT has also contributed significantly to graduate student training, the National Nuclear Physics Summer School, and undergraduate REU programs.

The DOE office of nuclear physics deserves praise for creating and adequately supporting the INT, and we strongly recommend that this support continue.

# 3. Recent Achievements

## 3.1. General considerations

Achievement in theory is often driven by experimental discovery. Correspondingly, experimental progress is often driven by compelling theoretical questions. One important role of theoretical effort is to capitalize on the success of existing experimental programs as well as to help guide them. Another important role for nuclear theory is to motivate and plan for new opportunities in an era where experimental facilities are ever more expensive and require extensive advance planning. Once new facilities begin to operate, theoretical support is essential in order to maximize their scientific return, just as increased theoretical support is essential to existing facilities. Whenever R&D funds are allocated to a new facility, like the Rare Isotope Accelerator (RIA) or the National Underground Science Laboratory (NUSL) it is crucial that a significant fraction of this money should support associated theory.

The existing facilities provide many case studies that support these statements. The Relativistic Heavy Ion Collider (RHIC) physics program arose from theoretical studies of the 1970s, which suggested that the state of the vacuum we live in could be changed under extreme conditions, such as high temperature or large baryon density. Theory played a crucial role in determining the required conditions, in identifying the parameters of a heavy ion collider under which these conditions can be achieved, and in specifying the required detector capabilities. In a similar fashion, theoretical ideas about the scale at which the quark-gluon substructure of hadrons becomes evident led to the construction of the Jefferson Lab's CEBAF facility, and the theoretical suggestion that neutrino flavor transformation takes place in the sun, as well as the theoretical work that went into the standard solar model, led to the design of the Sudbury Neutrino Observatory (SNO). Advances in the theory of nuclei at the limits of stability have contributed to the motivation for RIA. Many theoretical ideas, like neutrinoless double beta decay and tests of other physics beyond the Standard Model, are coming together in the drive toward NUSL, and new theoretical ideas about the structure of hadrons probed at high energy are providing the motivation for exploring the feasibility of an electron-ion collider.

Theoretical ideas and effort are as crucial to the success of these programs as they are to the identification of their primary objectives. Theory plays an essential role in analyzing and correlating the enormous wealth of data produced by RHIC and JLab. At RHIC, the sustained theoretical effort that followed the initial ideas has led to the development of analysis tools now used by the experimental collaborations. These tools are essential to the interpretation of new experimental findings, like the observed suppression of large transverse momentum pions.

Nuclear theory has made great strides since the time of the Koonin Report. We turn now to specific examples of recent theoretical achievement in the last seven or eight years. This list is not meant to be exhaustive but instead is illustrative of the breadth and the vitality of the field. It was formed by community input taken from a careful reading of the responses to our questionnaires.

## 3.2. Effective field theory and large-$N_c$ QCD in nuclear physics

Effective field theory (EFT) is a powerful framework based on controlled approximations to many problems with a natural separation of scales. In this formalism, the physics at "low" energies or long distances (chiral dynamics, for example) is represented by interactions constructed systematically, in the sense that all interactions allowed by fundamental symmetries are included. The physics at "high" energies or short distances is described by coupling constants that must either be calculated from a more fundamental theory (lattice QCD, for example) or must be fitted to a few measured quantities allowing many others to be predicted. These couplings, for example, summarize the incalculable and difficult to define short-distance part of the nucleon-nucleon potential. In the last few years, the EFT technique has been successfully applied to low-energy nuclear physics, yielding fundamentally new insight about physics that had been thought to be well understood, and providing promising new approaches to long-standing problems in the field.

The use of EFT concepts has led to significant development in our understanding of the two-nucleon system, and has provided analytic and systematically controlled calculations for astrophysically significant reactions, such as $n + p \to d + \gamma$ for big-bang nucleosynthesis and $p + p \to d + e^+ + \nu_e$ for stellar evolution.

EFTs that incorporate the chiral symmetry of QCD have guided the development of a systematic picture of the nucleon-nucleon and three-nucleon interaction and of isospin violating terms in the two-nucleon force. More recently, a number of model-independent results for three-body systems have been derived. EFT has also revitalized standard approaches to solving problems such as the nuclear shell model and nuclear density functional theory.

In the last few years, the effective field theory that exploits the light quark limit (the chiral expansion) and the limit in which the number of quark colors becomes large (large-$N_c$ limit) have been systematically and widely used to study properties of hadrons. For example, we have long known that simple quark models work well for hadron phenomenology, but have not understood how they arise from the fundamental QCD Lagrangian. It has been shown recently that the spin-flavor symmetries of baryons that are built into quark models emerge directly from QCD in the large-$N_c$ limit. This has led to controlled calculations of the spectra and properties of hadrons together with new understanding of the spin-flavor dependence of nucleon-nucleon interactions. Analyses done using the large $N_c$ limit have also clarified the role of the delta resonance in the chiral expansion. The combined large-$N_c$ and chiral expansion has been used to calculate many hadronic observables. Examples include strange quark form factors, electromagnetic polarizabilities, threshold electro- and photo-pion production, generalized Drell-Hearn-Gerasimov sum rules and parton distributions.

### 3.3. Lattice QCD

Lattice field theory has become an essential tool for calculating the properties of hadrons and of QCD thermodynamics directly from the QCD Lagrangian. As a result of a series of theoretical breakthroughs and the rapid rise of computing power, lattice QCD has begun to have a significant impact on the interpretation of experimental data and is providing new insight into nonperturbative strong interaction physics. Furthermore, essential computational tools have now been developed that promise quantitative evaluation of hadronic observables with the next generation of computers.

Lattice QCD calculations have provided valuable insights about the spectrum and structure of hadrons, the physics of chiral symmetry breaking and chiral restoration, and the QCD phase diagram and the equation of state of hot quark matter at temperatures up to a few times the critical temperature. Importantly for RHIC, lattice calculations have determined the temperature of the crossover or transition between hadronic matter and the quark-gluon plasma with a theoretical accuracy of better than 10%. Lattice QCD has also made it possible to calculate nonperturbative correlation functions, the static screening potential that serves as a phenomenological input to analyses of heavy quarkonia within a quark-gluon plasma, and the modifications of hadrons in the presence of a thermal medium. Turning to hadron structure, lattice calculations have determined the string potential between static quarks and how this changes if the string is excited, demonstrated string breaking at large separation, and explicitly shown the presence of topological excitations. Methodology has been established to calculate many observables directly relevant to experiments at JLab, RHIC-Spin, and Bates, including nucleon electromagnetic and strangeness form factors, contributions of the quark spin and quark angular momentum to the nucleon spin, moments of parton distributions as well as generalized parton distributions, and nucleon to delta transition form factors. Although not yet completely quantitative, current lattice calculations of glueballs, exotic mesons, excited nucleons, the H particle and the recently discovered pentaquark can guide the search for, and help understand the properties of, these exotic particles.

A crucial algorithmic breakthrough has been the development of lattice techniques which implement chiral symmetry exactly on the lattice by the use of perfect actions, domain wall fermions, or overlap fermions. These have led to credible calculations of kaon observables in which chiral physics plays a crucial rule, such as the direct CP-violating parameter, and will play a similar role for light hadron properties in the next generation calculations on larger lattices using lighter quark masses. Recent advances in the development of lattice variants of chiral perturbation theory are also crucial since they facili-tate reliable extrapolation to the continuum, to light quark mass, and to the large volume limit.

## 3.4. Generalized parton distributions and nucleon spin structure

The traditional way to probe the quark and gluon structure of the nucleon is to measure either the elastic form factors, which provide the spatial distributions of the quarks, or the parton distributions, which are the momentum-space distributions of the quarks and gluons. Recently, new nucleon observables called generalized parton distributions (GPDs) have been discovered which provide the joint quantum phase-space (position and momentum) distributions of the partons. These generalized quantities include the long-known form factors and parton distributions as limiting cases, but go far beyond. If measured, they can be used to make a complete mapping of the quarks and gluons in the nucleon. This information is extremely powerful since, for instance, it allows one to finally understand the decomposition of the nucleon spin into contributions from the spin and motion of quarks, and from the gluons.

These distributions have driven the development of a new class of experiments through which they can be measured, such as deeply-virtual Compton scattering and hard exclusive meson production. It has been in shown in perturbative QCD that such processes are factorizable in the sense that initial and final state effects can be systematically controlled, allowing the extraction of the desired distributions. The experiments from H1, ZEUS, HERMES, and CLAS collaborations have verified the ability to extract GPDs from these processes, and an extensive program for the JLab 12 GeV upgrade and for a proposed electron-ion collider has been formulated.

## 3.5. *Ab initio* calculations of light nuclei

Recent advances have, for the first time, made possible direct calculations of the structure of nuclei up to mass 12 directly from the free-space interactions between nucleons. Nucleon-nucleon interactions obtained from scattering data, when combined with a modern three-nucleon interaction, can accurately reproduce the binding energies and spectra of light nuclei. The isospin dependence of the three-nucleon interaction is crucial to the quantitative understanding of light nuclear spectra and, when combined with microscopic models of electroweak currents, makes it possible to accurately describe nuclear form factors, response, and decays.

Rapid progress has also been made in calculations of reactions in light nuclei. For example, scattering calculations have made it possible for polarized $^3$He to be used as a reliable probe of neutron properties. Moreover, the uncertainties in crucial astrophysical reactions including pp capture and neutrino induced deuteron breakup have been reduced. Conventional and hybrid EFT approaches have limited the theoretical uncertainties in the pp capture rate to less than 1%, and have drastically reduced the errors in the breakup reaction crucial to SNO. Such reactions are a key ingredient for tests of the standard solar model, whose validity has been brilliantly confirmed by the recent SNO measurements.

### 3.6. Theory of complex and exotic nuclei

Large-scale shell-model calculations and modern mean-field approaches with accurate effective interactions have provided a much-improved microscopic description of fp-shell nuclei, such as doubly-magic $^{56}$Ni. Those nuclei are of particular interest as they are amenable to different theoretical treatments while studying the competition between single-particle and collective excitations, including superdeformed bands. Another major theoretical achievement has been the development of new theoretical tools, such as the Shell Model with Monte-Carlo techniques and the nonrelativistic and relativistic energy density functional theory capable of properly treating pairing correlations in weakly bound nuclei. Microscopic calculations using the new tools, supported by early radio-active ion beam experimentation, have shown that many traditional concepts of nuclear structure break down or need to be revised in drip-line systems, wherein many-body correlations govern static and dynamic properties, i.e., determine nuclear stability, the excitation pattern, and decay. The predicted changes of shell structure (quenching of traditional magic gaps and appearance of new ones) together with the presence of low-energy skin excitations can have a decisive effect on r-process abundances.

Improved theoretical descriptions of reactions with fast radioactive beams have enabled the extraction of subtle structural information on wave functions of weakly-bound, neutron-rich halo systems. On the proton-rich side of the valley of stability, a detailed understanding of proton emitters has been achieved; these are prime examples leading toward the ultimate goal of a unification of nuclear structure with reaction theory.

These developments in nuclear structure theory have been crucial to the scientific justification for the emerging field of radioactive ion beams worldwide, in particular for the RIA proposal in the US.

### 3.7. Diagnostics of hot and dense matter

Recent progress in the theoretical understanding of diagnostic probes of the hot and dense matter produced in high-energy heavy-ion collisions has had a significant impact on the RHIC program. Parton energy loss and the resulting jet quenching in a dense medium have been studied using perturbative QCD calculations that incorporate important interference effects. This provides the theoretical underpinning for a systematic experimental program utilizing hard probes, such as jets and their hadronic fragments, for the investigation of the hot and dense matter immediately after its creation. These developments are built upon a systematic understanding of the "hard thermal loop" effective theory for the quasiparticles of the quark-gluon plasma at high temperature. This has been developed into a powerful tool for the prediction of other diagnostic probes also, for example providing predictions for the emission of photons. Recent advances have enabled the calculation of plasma transport coefficients to leading order in the interaction strength, providing the basis for improved transport calculations.

Another major theoretical achievement is the development of new tools, such as multi-particle correlations resulting from anisotropic collective flow patterns, for the study of the thermodynamic properties of the created matter. Three-dimensional solutions of relativistic fluid dynamics have linked the initial conditions, energy density and pressure, at which hot matter is created to the measured azimuthal anisotropy of the collisions. The comparison between the data and these calculations has provided compelling evidence for early thermalization in collisions at RHIC.

## 3.8. Classical gluon dynamics and high energy particle production

Significant conceptual progress has been made over the past decade on one of the most intractable problems of quantum chromodynamics: the high-energy, as opposed to short-distance, limit of strong interaction phenomena. The fundamental new insight is that a semiclassical description of color fields is appropriate, when a hadron or nucleus is probed at very high energy but low virtuality, and the probe coherently interacts with many gluons at once. In the extreme high-energy limit, which is reached earlier for large nuclei than for individual hadrons, the system is described by a superposition of randomly oriented classical color fields (the "color glass condensate"). This novel quasi-classical state, whose properties are the same for all hadrons, is weakly coupled, meaning that quantum corrections can be calculated perturbatively. The entropy released by the decoherence of such quasi-classical color fields is thought to be responsible for a large part of the particle multiplicity measured in the collisions of nuclei at RHIC.

## 3.9. Color superconducting phases of cold dense quark matter

Recent theoretical advances have shown for the first time that QCD provides rigorous analytical answers, with no unresolved nonperturbative gaps in understanding, to the question: "what are the properties of matter squeezed to arbitrarily high density?" It has long been known that cold dense quark matter, as may occur at the center of neutron stars, must be a color superconductor. Recent theoretical effort has made this subject both more quantitative and much richer. An ab-initio calculation of the pairing gap and critical temperature at very high densities has now been done, and the properties of the phase at these densities have been determined. The material is a color superconductor but admits a massless "photon" and behaves as a transparent insulator; the material is a superfluid with spontaneously broken chiral symmetry. At densities that are lower but still above that of deconfinement, color superconducting quark matter may in a particular sense be crystalline. The analog of this phase of quark matter should be accessible in experiments in which trapped ultracold fermionic atoms in two different hyperfine states interact, forming a crystalline condensate.

## 3.10. Theoretical studies of type II supernovae

Developments in many areas of physics have been synthesized in core collapse supernovae simulations that now include realistic physics assumptions and microscopic nuclear physics input. This has required new calculations of neutrino opacities and weak interaction rates from the nuclear theory community, and advances in other areas like the development of general relativistic Boltzmann neutrino transport. Such progress has made it possible to isolate the effects of variation in different physics inputs, for example the nuclear equation of state, on the development of the supernova shock.

Recent theoretical effort has also yielded significant advances in our understanding of nucleosynthesis in core-collapse supernova. We have progressed from the initial suggestion that the neutrino-driven wind may be viable for r-process nucleosynthesis to the point where we now understand how various parameters characterizing this environment influence abundance patterns. Also, a new type of nucleosynthesis called the "neutrino process" has recently been proposed, arising from neutrino-induced neutron spallation in the outer layers of a supernova. This process plays a significant role in determining the abundances of certain rare isotopes.

## 3.11. Confirmation of the standard solar model

The standard solar model has been resoundingly confirmed by measurements made at SNO, completing the picture that had emerged from measurements made at other neutrino observatories. This scientific success story would not have been possible without recent nuclear theory calculations of key cross sections and screening corrections and neutrino oscillations. For example, recent calculations have crucially changed the predicted rate at which $^8$B and the high energy neutrinos to which SNO and Super-Kamiokande are sensitive are produced. The role of nuclear theory goes beyond its input to the standard solar model predictions, however. When SNO was originally proposed, the neutrino-deuteron cross section was sufficiently uncertain that it was assumed impossible to accurately measure the total neutrino flux. Dedicated theoretical effort in the intervening years determined the required cross section, allowing SNO to make its flux measurement and in so doing verify that the sun produces the neutrinos at the rate predicted by the standard solar model as well as demonstrate their oscillation.

## 3.12. Interdisciplinary many-body physics

Nuclear physicists have a history of making seminal contributions to studies of strongly correlated quantum many-body systems. Nuclear theorists have been very successful in exploiting the resulting ties to a wider physics community, both contributing to achievements in other disciplines and exploiting them within nuclear physics. Important examples of areas where nuclear theory has had such impact include cold atomic gases, conductance in mesoscopic systems, and low-energy tests of QCD in atomic systems.

Nuclear physicists have recently played an important role in advancing our understanding of cold trapped atomic gases. The resulting theories have expanded our knowledge of the dynamics of condensate formation, and the ground state of dilute gases of bosons and fermions. In addition, weakly rotating atomic gases and the formation of vortices have been studied extensively by nuclear physicists. All of this work has played a key role in explaining important experimental observations as well as in motivating many of these beautiful experiments.

Even in cases where the underlying interactions are very simple, very complex behavior can arise in quantum systems. The conductance and general transport properties of quantum dots and nanostructures are crucial observables which can sometimes be interpreted in terms of scattering processes. Random matrix theory, originally developed to describe nuclear level densities, has recently helped to understand the statistical properties of the conductance of quantum dots, yielding new insight into the mesoscopic fluctuations. It has also been used to describe spectral properties in both nuclear and condensed matter systems.

Many-body theory is also a crucial tool for performing low-energy tests of theories whose fundamental scales are higher in energy. For example, in order to confront precise atomic physics parity violation experiments with QCD, equally precise theoretical calculations are required. This goal has been achieved recently by several groups, who have performed calculations for atoms in the alkali region which are accurate to better than 1%. The result has been that such experiments are now competitive with other methods as a testing ground for QCD and as a probe for physics beyond the Standard Model.

# 4. Opportunities

There are compelling scientific opportunities that argue strongly for significant increases in the support of nuclear theory in the United States. Some of these opportunities are described in the following paragraphs. While input from the nuclear theory community was essential in formulating this list of opportunities, it must be emphasized that the opportunities described herein constitute a representative collection rather than an exhaustive one. They serve to illustrate the breadth, depth and vibrancy of this multi-faceted field. In addition, new and exciting opportunities in nuclear theory, impossible to envision at the time of writing of this report, will undoubtedly arise in the long term, and even in the short term. It is therefore of paramount importance that support for future efforts in the field be sufficiently flexible that such new, unforeseen opportunities can be seized and brought to fruition in a timely fashion.

## 4.1. Lattice QCD

QCD is an integral part of the Standard Model and is universally accepted as the fundamental theory of strong interactions. The QCD vacuum and its hadronic excitations are intricate quantum mechanical systems composed of strongly coupled, ultrarelativistic quarks and gluons. Lattice field theory is the only method at present to solve, rather than model, nonperturbative QCD, providing the opportunity to understand the physical mechanisms of confinement and chiral symmetry breaking, to calculate hadron properties and hadron-hadron interactions, and to map out the phase diagrams of strongly interacting matter.

There are a number of compelling reasons why now is the right time for a major investment in lattice QCD by the US nuclear community. Years of effort in lattice field theory have now developed the algorithms and theoretical tools for definitive calculations of many important hadronic observables. These calculations require tens of Teraflops, and for the first time cost-optimized computers for lattice QCD are available at the cost of $1M per sustained Teraflops. Furthermore, successful nuclear physics programs at Jefferson Lab and RHIC urgently need to make connections to QCD, and a dedicated effort in lattice field theory will help to realize the full potential of these large-scale facilities.

Quantitative solution of QCD on a lattice requires including the effect of dynamical quarks in the Dirac sea and controlled extrapolation to small lattice spacing, to large lattice volume, and to physical light quark mass. While some important lattice calculations will still be beyond reach in the next 5 years, much first-rate physics requires resources estimated to be in the range of 10-15 Teraflops. We sketch these physics opportunities here, and describe the initiative motivated by these opportunities in Section 5.5.

### 4.1.1. *Understanding the QCD vacuum*

The QCD vacuum is characterized by two crucial nonperturbative phenomena: chiral symmetry breaking and confinement. A combination of analytical methods and lattice theory has demonstrated that the first of these can be understood as originating from the instanton-induced interaction between light quarks, but the "structure" within the QCD vacuum that is dominantly responsible for the mass gap in QCD, and hence for confinement, remains elusive. Investment in the lattice effort will enable a systematic investigation of mechanisms for chiral symmetry breaking and color confinement.

### 4.1.2. *Hadron spectrum and hadronic structure*

Current computational resources limit lattice calculations with dynamical fermions to a box sufficiently small that the pion mass must exceed 500 MeV, forcing theorists to temporarily work in an artificial world in which the pion cloud of a nucleon is strongly suppressed and thereby producing discrepancies with experiment that can range from 10% to 50%. An appropriate lattice investment will enable calculations in the chiral regime, which includes the pion cloud, with the prospect of calculating many hadronic observables at the few-percent level. The spectrum of the lowest mass hadrons with distinct quantum numbers may be calculated particularly reliably, providing the opportunity of calculating exotic mesons, excited nucleons, and states like the pentaquark that lie at the heart of frontier searches for excited and exotic hadrons at JLab.

Experiments at JLab, Bates, RHIC-Spin and elsewhere provide rich and precise measurements of the quark and gluon structure of the nucleon. An investment in lattice QCD opens the way to precise lattice calculations of nucleon electric and magnetic form factors, strangeness form factors, the fraction of the nucleon spin arising from quark spin and quark angular momentum, nucleon polarizabilities, the momentum dependence of generalized parton distributions, and the nucleon to delta transition form factors indicative of baryon deformation, all of which can be compared with contemporary or planned experiments.

### *4.1.3. Hot and dense quark matter*

Current lattice calculations have determined the energy density and pressure of the quark-gluon plasma at zero baryon chemical potential with an accuracy of order 20%. A lattice investment will allow calculations on larger lattices with more sophisticated actions, reducing the theoretical errors to of order a few percent and allowing a thorough study of the nature of the phase transition as a function of the masses of light and strange quarks. Recent developments also allow these calculations to be extended to baryon densities that are nonzero but not too large, including the regime relevant for heavy ion collision experiments. An important goal is to refine these calculations to locate the second order critical point in the QCD phase diagram precisely.

There are also challenging problems that require new approaches and algorithms. For example, it would be of great interest to use lattice techniques to study the novel color superconducting phases of QCD and the equation of state at high baryon density, where conventional algorithms break down completely because of the fermion sign problem. Theorists have developed new methods, such as the meron cluster algorithm, which hold promise but have to date solved this challenging problem only in limited cases. These ideas need to be further generalized and developed before they can be exploited. Success here cannot be guaranteed, but if these challenges can be overcome the result would have major impact in condensed matter physics, in addition to nuclear physics. Similarly, techniques are not fully developed to go beyond thermodynamics and study real time physics like transport properties and dilepton emission, but maximum entropy methods have promise. With the new talent and theoretical effort that the initiative we describe in Section 5.5 would engender, one can also expect new ideas and approaches to continue to expand the scientific opportunities for using lattice QCD.

## 4.2. JLab and RHIC Phenomenology

The two flagship facilities of the Department of Energy's nuclear physics program, Jefferson Laboratory and the Relativistic Heavy Ion Collider, were constructed to provide answers to fundamental questions like "How does QCD give rise to hadrons and their properties?" and "What are the properties of the quark-gluon plasma?" Understanding the QCD predictions for these questions requires effort in lattice QCD and effective field theory, described elsewhere. Understanding the wealth of new data now coming from these facilities, and using these data to test predictions of QCD, requires sustained theo-retical effort of a different character. Moreover, such an effort is crucial if the investment in these facilities is to provide maximal and optimal return. We give several concrete examples of opportunities in this direction here. Once RIA and NUSL are making measurements, theoretical opportunities of this character will arise from these facilities also.

*4.2.1. Hadron structure and spectroscopy*

The theoretical description of the measurements of various observables of multiparticle final states in Jefferson Lab meson production experiments poses a serious challenge for theorists. The description of these multiparticle final states, including the constraints of unitarity and analyticity, as well as the treatment of many coupled channels, pushes reaction theory for these kinds of processes into uncharted territory. The high quality of the data that are being obtained means that theoretical efforts have to focus on many previously ignored details, to produce descriptions of the production processes commensurate with the experimental precision. Furthermore, analysis and understanding of the data taken in the planned program of meson production at higher energies will require theoretical tools developed in the next few years, before the upgrade to JLab is completed. These tools are also appropriate for the hadronic physics program of CLEO-c. An effort of this nature is a prerequisite to making contact between hadronic data and effective field theories like chiral perturbation theory, in order to test the range of applicability of the latter. To take full advantage of this opportunity, a focused reaction theory effort is crucial.

Progress in reaction theory is a prerequisite to obtaining information on hadronic spectroscopy, which is in turn vital input into the quest to understand QCD at a nonperturbative level. Spin degrees of freedom are increasingly being exploited to probe hadron structure and reaction mechanisms at lower energies, demanding an improved understanding of the nonperturbative dynamics of quark and gluon spin in QCD. One of the goals of lattice QCD described in 4.1 is the ab-initio calculation of the properties of the lightest hadrons with distinct quantum numbers, an opportunity that requires progress in reaction theory in order to compare with data. Models of hadrons and their properties serve to integrate a vast range of information about phenomena and higher excitations to lowest order, within relatively simple frameworks. Such models have a long history of providing the kinds of understanding and qualitative insights that are the prerequisites to more rigorous analyses. The interplay between lattice simulations, the predictions of effective field theory approaches, large-$N_c$ approximations, calculations of higher twist effects and the role of parton transverse and orbital motion, models that include dynamical quarks and gluons, and reaction theory will, in the face of the expected high precision data from JLab and its 12 GeV upgrade, help us to make progress in our understanding of nonperturbative QCD. Achieving this goal requires carrying out effective and timely research on all the components of this effort.

*4.2.2. RHIC-Spin phenomenology*

The goal of the RHIC-Spin program is to advance our understanding of the spin structure of the nucleon by disentangling the contribution of the gluons, quarks, and antiquarks. It will also be possible to get a first look at the quark transverse spin distribution in the nucleon, which is sensitive to relativistic spin dynamics. Various polarization asymmetries (both single- and double-spin asymmetries) will be measured. Achieving the goal of extracting the polarized parton distribution functions from the asymmetries requires a sustained theoretical effort in perturbative QCD. The theory of polarized quark and gluon distributions in the nucleon has advanced dramatically in recent years. Future progress will require collaboration between theory and experiment as the data comes in. The theoretical component of this effort is crucial to the RHIC-Spin program.

*4.2.3. Quantitative analysis of measurements of QGP observables*

The results from the first RHIC runs have revealed new phenomena, which clearly probe the properties of the highly excited matter produced in nuclear collisions. The critical role of (nuclear) theory will be to make a quantitative connection between the increasingly precise data and the physical properties of the hot QCD matter which are responsible for the observed effects. Making this connection will require more precise calculations of the relevant diagnostic signatures, and these calculations must be solidly based on QCD. For instance, the current RHIC discoveries motivate the sustained theoretical effort required to extract quantitative estimates of the stopping power, opacity, shape and evolution of the dense medium from the ensemble of jet quenching data; to obtain solutions of 3-dimensional hydrodynamics using the equation of state calculated on the lattice and including effects of dissipation; and to carry out parton transport calculations with realistic initial conditions and consistent treatment of gluon radiation. Doing the baseline calculations needed to attain a precise understanding of the implications of precise data will require sustained collaboration among theorists with different skills and expertise. If the soft collinear effective theory (SCET, see Section 4.5.4) can be extended to include effects of a dense medium, it might permit a more unified description of hard probes of the quark-gluon plasma.

*4.2.4. Understanding thermalization and mean-free paths in QCD matter*

Data indicate that the classical initial state in a heavy ion collision thermalizes more quickly than qualitative theoretical estimates had indicated. The basic ingredients of an infrared-safe quark-gluon transport theory have recently been formulated, but a quantitative treatment of thermalization requires systematic study of the dissipative dynamics of color fields, formulation of the transition from the classical fields to the parton transport domain, and numerical solutions of the transport equations. Data also indicate that mean free paths are short, not much longer than the spacing between partons. This makes it unlikely that any perturbative calculation of the shear viscosity or the opacity will be quantitatively reliable at RHIC. Theorists have recently used methods developed in string theory to obtain insight and nonperturbative answers for these quantities in supersymmetric QCD, and are working toward a calculation in real-world QCD.

## 4.3. Nuclear Structure Theory

Nuclear structure theory deals with fundamental questions permeating modern science, such as "How do complex systems emerge from simple ingredients?" and "How do simplicities and regularities in complex systems emerge?" These themes complement each other beautifully. Indeed, the goal of nuclear structure theory is to build the unified, comprehensive microscopic framework in which bulk nuclear properties (including masses, radii, and moments, structure of nuclear matter), nuclear excitations (including the variety of collective phenomena), and nuclear reactions can all be described. Microscopic theory also provides the solid foundation for phenomenological models and coupling schemes which have been applied so successfully to explain specific nuclear properties, including elegant simplicities exhibited by collective modes appearing in strongly interacting nuclear many-body systems.

Exotic short-lived nuclei are the critical new focus in this quest. The extreme isospin of these nuclei and their weak binding bring new phenomena to the fore and isolate and amplify important features of the nuclear many-body problem. This new arena is therefore key to building a unified theoretical foundation for understanding the nucleus in all its manifestations - from the stable nuclei that exist around us to the most exotic nuclei, and even to exotic forms of nucleonic matter which exist, e.g., in neutron stars. RIA and other exotic beam facilities allow unique insights into the quantum many-body nature of nuclei by providing access to their most extreme manifestations and by allowing precise control of the number of bodies in these systems. As we describe below, recent theoretical and experimental achievements, coupled with the experimental discoveries that RIA would provide, are focusing new attention on a host of unsolved issues in nuclear structure and offer unprecedented opportunities for the next decade and beyond.

### 4.3.1. Microscopic nuclear structure theory

Projecting from the recent successes in theoretical nuclear structure studies for light and medium-mass nuclei, one can expect significant advances in the field. Ab-initio calculations, based on realistic two- and three-nucleon interactions, will reach nuclei with $A \approx 16$ and will be extended to neutron-rich systems. Progress in larger nuclei will be achieved by optimizing the enormous configuration spaces using novel quantum many-body techniques (such as Monte-Carlo or Density Matrix Renormalization Group methods). Parallel with this development, recent progress in the area of effective interactions offers the hope for constructing microscopic effective forces which are consistent with these effective operators. EFT methods incorporating the chiral symmetry of QCD are important in constraining nuclear interactions and may also prove valuable in the wider context of many-body approaches.

### 4.3.2. Nucleon-nucleon effective field theory

The application of rigorous chiral methods to the NN system has been extremely productive, enabling an understanding of the three-body interactions as well as successful representation of the low-energy NN data within a consistent power-counting approach.

At the present time two- and three-body systems such as the deuteron and triton are under control, but future progress of such methods must involve the understanding of four-body and higher body interactions as well as the application of such methods to $^4$He and heavier systems. Preliminary work in this regard is underway but much work remains.

### 4.3.3. *Toward the universal microscopic energy density functional*

For heavy nuclei, a critical challenge is the quest for the universal energy density functional, which will describe properties of finite nuclei as well as extended asymmetric nucleonic matter, as found in neutron stars. This quest is driven by new data of both terrestrial and cosmic origins, especially on nuclei far from stability, where new features, such as weak binding and altered interactions, make extrapolations of existing models (inspired by data on known nuclei) very unreliable.

While modern nuclear energy density functionals predict nuclear binding energies of known nuclei with an impressive rms error of less than 700 keV, developing a universal approach will require a better understanding of the isovector and density dependence of the functional as well as an improved treatment of many-body correlations. Studies of nuclear matter at very low density near the drip lines, where a transition to non-uniform phases takes place, and at higher temperatures and isospin (nuclear matter equation of state relevant to supernovae) are particularly interesting in this context. It is also crucial to extend such theories to reliably predict weak reaction rates in these new exotic regimes.

### 4.3.4. *Coupling of nuclear structure and reaction theory*

Tying nuclear structure directly to nuclear reactions within a coherent framework applicable throughout the nuclear landscape is an important goal. For light nuclei, ab-initio methods hold the promise of direct calculation of low-energy scattering processes, including those important in nuclear astrophysics, and tests of fundamental symmetries. In nuclear structure for heavier nuclei, the continuum shell model and modern mean-field theories allow for the consistent treatment of open channels, thus linking the description of bound and unbound nuclear states and direct reactions. On the reaction side, better treatment of nuclear structure aspects is equally crucial. The battleground in this task is the newly opening territory of weakly bound nuclei where the structure and reaction aspects are interwoven and where interpretation of future data will require advances in understanding of the reaction mechanism.

### 4.3.5. *Complex nuclei: Nuclear dynamics and symmetries*

Microscopic understanding of nuclear collective dynamics is a long-term goal. Large amplitude collective motion, such as that seen in fission, fusion, shape coexistence, and nuclear phase transitions, provides a particularly important challenge. Some of these phenomena are manifestations of many-body quantum tunneling, an intriguing mechanism arising in a multitude of physical contexts.

Advances in many-body theory (e.g., the imaginary time method, the generator coordinate method, projection operator techniques), powerful numerical algorithms, and significantly improved computational resources promise progress in the field of nuclear dynamics, where high-resolution data offer new theoretical challenges. Examples abound in such areas as the spectroscopy of the heaviest elements, new kinds of nuclear deformations associated with nucleonic spins and currents, spectacular cases of phase-transitional behavior, and coupling between coexisting nuclear states. The microscopic description of the spontaneous fission problem – one of the oldest but still not fully understood nuclear decay modes – is on the horizon. This, together with the identification of new structural symmetries and coupling schemes, will give insights into the astonishing simplicities and regularities that complex nuclei exhibit.

## 4.4. Nuclear Astrophysics

The melding of nuclear physics and astrophysics plays a central role in understanding the evolution of stars, galaxies and the universe, as well as the origin of all the elements which make up our world. Nuclear processes are an integral part of the stellar interior. They determine, in a cosmic battle with gravity, the evolution of stellar objects, the violent explosions of massive stars, and the synthesis of the elements, hence the appearance of life itself. To understand large scale astrophysical phenomena, theoretical nuclear physics on microscopic scales is vital.

Opportunities in nuclear astrophysics arise from the discovery potential of new observatories, such as X-ray telescopes, from the exploration of the rich astrophysical phenomenology of neutrino oscillations, and from the drive to understand extraordinarily dense stars and spectacular phenomena like gamma ray bursts, in which nuclear physics plays a crucial role. In addition, the synthesis of and progress in the many physics aspects of core collapse supernova is leading us to construct quantitative models of these objects.

*4.4.1. Quantitative understanding of supernovae*

We can expect that advances in many areas of physics and in computer technology (petascale computing) will combine to finally give realistic three-dimensional core collapse supernova models that correctly describe explosions as seen by astronomers and make precise predictions for future neutrino signals in underground detectors. In the near future, with terascale computers, realistic physics can be included in 2-dimensional models. The success of this ambitious and multidisciplinary program requires developments in neutrino transport and magneto-hydrodynamics as well as improved microscopic nuclear physics input. Weak interaction processes play a decisive role in the early stage of the core collapse, which defines the ensuing explosion. This requires better theoretical modeling of both electron capture and beta decay of short-lived neutron-rich nuclei as well as the cross sections for neutrino-nucleon scattering.

Another area where improved nuclear physics input is needed is the nuclear matter equation of state, a crucial input for the core collapse calculations. Studies of neutrino oscilla-tions in supernovae are essential in order to understand any supernova neutrino signal that will be seen with existing or future detectors. Although atmospheric and solar neutrino results imply mixing in the outer layers, mixing in the interior cannot be ruled out. This will affect the explosion dynamics, nucleosynthesis and neutrino signatures. If a phase transition to quark matter occurs within the neutron star newly-born in a supernova, this could also affect neutrino signatures. Theoretical effort will go hand in hand with experimental advances in the field. In addition to the neutrino detectors at SNO and Super-Kamiokande, LIGO is now on line and looking for supernovae within our galaxy. The future RIA and NUSL facilities will provide basic input to weak interaction physics that will help to calibrate theoretical models.

### 4.4.2. Nuclear physics of gamma-ray bursts

Although gamma-ray bursts have been observed for several decades, it is only in the last few years that observational evidence concerning their origin has been pouring in. This has allowed dramatic theoretical advances, with evidence pointing to rapidly rotating core collapse events called 'collapsars' or 'hypernovae' as the origin of the longer duration bursts and neutron star mergers or neutron star-black hole mergers as the origin of short bursts. We are beginning to understand these environments in more detail and in the process are mapping out the weak interaction physics, hydrodynamics, and nucleosynthesis that occurs during gamma-ray bursts.

### 4.4.3. Neutrino astrophysics

The near future will no doubt produce a large set of new neutrino physics observations from experiments such as Mini-BooNE, MINOS and others. A major challenge will be to understand the implications of these discoveries for solar, supernova and gamma-ray burst physics as well as for big-bang nucleosynthesis, including connections with the r-process and with future RIA measurements.

### 4.4.4. Neutron star structure

New insight into neutron star physics will come from new observations in X-ray astronomy and, with luck, from the detection of neutrinos from a supernova or the observation of gravity waves from a compact star merger. It is a major theoretical challenge to understand the implications of such new data for the equation of state and the composition of dense nuclear matter (for example, does it include hyperons?). Furthermore, the combination of observed properties of neutron stars and careful theoretical analysis of the influence of a color superconducting quark matter core will eliminate some of the possibilities that theorists have proposed for the cold but dense regime of the QCD phase diagram and could yield a compelling case for or against the existence of quark matter cores within neutron stars.

## 4.5. Fundamental Symmetries and Beyond the Standard Model

In order to cleanly separate hadronic physics, parameters of the electroweak standard model, and effects that arise from physics beyond the standard model, one requires methods which permit accurate predictions for processes involving hadrons to be made from the underlying elementary interactions. Methods based on symmetry offer some of the most powerful means by which this can be accomplished. There exist many challenging opportunities in this regard, and below we outline some of them. The recent NSAC report on fundamental physics with neutrons describes experimental opportunities that correspond to the theoretical opportunities we outline in 4.5.1 and 4.5.4.

### 4.5.1. Baryon asymmetry and electric dipole moments

Two major enigmas of contemporary physics are (i) Why is there such a strong asymmetry between baryons and antibaryons in the universe? and (ii) What is the mechanism which produces CP violation? Actually, these two questions are related, since a fundamental theorem shows that (i) can only come about in the presence of CP violation (together with non-equilibrium conditions). Recent and future experiments probing for electric dipole moments of elementary particles, nuclear, and atomic systems will provide increasingly tight limits (or even an actual measurement) of such electric dipole moments, and hence of CP violation. It is a major theoretical challenge to understand the restrictions which this will place on the corresponding baryon number asymmetry.

### 4.5.2. Beta decay

The most precise probe of the CKM matrix element $V_{ud}$ is provided by nuclear and neutron beta decay. However, despite the impressive accuracy of such measurements an important limitation in the analysis of such experiments is the radiative correction, for which uncertainties exist in the hadronic component as well as in the higher order terms in $Z^n \alpha^m$ in the nuclear case. Such corrections need to be evaluated, in order for the full potential of such measurements to be reached.

The only probe for the existence of a Majorana component of the neutrino mass is through the neutrinoless double beta decay process. Such methods have the potential to limit such a quantity to the fraction of an electron volt level, and the experiments which have this reach are at hand. What is needed to achieve this goal, however, are parallel nuclear matrix element calculations at the requisite level of precision. Present structure calculations have considerable uncertainty and need to be significantly improved if this goal is to be achieved.

### 4.5.3. Quantitative determination of (beyond) Standard Model parameters

In their quest to understand the dynamics of non-perturbative QCD in hadrons, nuclear theorists can and will play a significant role in the extraction of fundamental parameters of the electroweak standard model like $V_{ub}$, $V_{ud}$ and $\varepsilon'/\varepsilon$, crucial to our understanding of fundamental symmetries. Uncertainties in the extraction of such quantities are dominated by the uncertainties in theoretical quantities entering the extractions, primarily from hadronic matrix elements of operators. Analogously, uncertainty in our knowledge of hadronic matrix elements and hadronic radiative corrections is the primary limiting factor in the use of rare meson decays and other precision experiments to learn about new physics beyond the standard model.

In order to interpret any experimental signals of physics beyond standard model at a quantitative level, it is essential to be able to compare with precise radiative correction calculations within a range of scenarios and a major limitation in this regard is the reliable evaluation of hadronic effects. The importance as well as the difficulty of such estimation was evidenced, for example, in the analysis of the recent muon (g-2) experiment, but similar difficulties arise in the analysis of both Fermi and neutron beta decay. The use of realistic models or of chiral methods should be able to significantly reduce these particular uncertainties. This illustrates the general point that research in nuclear theory can have significant impact in our understanding of the electroweak standard model, and ultimately of whatever more complete theory subsumes it.

### 4.5.4. *Soft collinear effective theory*

Soft collinear effective theory (SCET) is an important new tool which provides a unified description of hard processes involving light partons in QCD, relying crucially on all the symmetries of QCD. All previous formulations of effective field theory have only been applicable to processes in which hadrons are either heavy or soft. SCET brings rigorous theoretical control to a whole new regime of reactions involving energetic hadrons formed from energetic, collinear quarks and gluons. Examples of future applications include the analysis of form factors being measured at JLAB, deeply virtual Compton scattering, the Drell-Yan process, and many other reactions in which energetic hadrons are produced. SCET will also play an important role in the extraction of fundamental parameters of the electroweak theory from decays of heavy mesons, as it can be used to analyze purely hadronic decays with energetic hadrons in the final state. SCET organizes previously intractable calculations, cleanly separating ("factorizing") perturbative and nonperturbative contributions in the strong coupling constant. It provides new methods for deriving QCD factorization theorems and performing all-order perturbative resummations, and provides a systematic method for analyzing the nonperturbative corrections to factorization, which are suppressed by ratios of energy scales.

### 4.5.5. *Hadronic parity violation*

Parity violating signals in $\Delta S=0$ hadronic reactions have been studied for nearly half a century but a consistent picture has yet to emerge. There are a number of reasons for this situation but certainly one of the most important is that this is a nonperturbative, strongly-interacting hadronic system. Past work has been framed within a meson-exchange approach, but recent analyses have been EFT-based and, combined with new experiments over the next few years, should lead to a reliable extraction of the coefficients which characterize the short distance physics. The extractions themselves will necessitate close collaborations between theorists and experimentalists, and further interpretation of the results in terms of QCD implications will require extensive theoretical analysis.

## 4.6. Interconnections: Interdisciplinary many-body physics

Phenomena observed in nuclear physics are mirrored in a host of quantum systems, and nuclear theorists have played a leading role in exploring and understanding them. Single-particle structure and pairing are ubiquitous quantum phenomena which are understood within many regimes where one or the other dominates. Examples include studies of shell structure in liquid helium drops and pairing and superconductivity in small metallic grains. Problems with more than one significant scale, though, remain an important challenge. For example, one would like to understand systems where the single-particle level spacing and pairing gaps are of the same order, as well as the transition between the small and large pairing-gap regimes.

Large scale quantum simulations will continue to play a significant role in developing our understanding of these strongly correlated quantum many-body systems. Nuclear theorists have advanced the state of the art in simulations of strongly-interacting fermions, for both lattice formulations of condensed matter and quantum field theories, and for continuum problems in nuclear and condensed matter physics. In some specific cases exact formulations free of the fermion sign problem have been devised. In others very accurate, approximate methods have been developed. These ideas must be fully developed in order to realize their full promise. Advances in this area would dramatically affect all areas of physics.

In the future, nuclear theory will be even more closely tied to other physics disciplines. Experiments in atomic gases, and in mesoscopic systems more generally, are rapidly advancing, allowing direct tests of theories of strongly correlated Fermi systems. In some cases these experiments can provide quite direct tests of phenomena not terrestrially observable at the nuclear scale. Nuclear theorists can play a leading role in understanding such experiments and advancing the theory of quantum many-body systems.

# 5. Recommendations and Initiatives

## 5.1. Guiding principles and aims

In order to most effectively support the scientific program described in the NSAC Long Range Plan for Nuclear Science, the community of theoretical nuclear physicists in the United States must attract and retain the best and most talented young scientists. The funding agencies can help in many ways in this endeavor, and they can enhance the effectiveness of the nuclear theory community by providing adequate means of research support. In return, the agencies can expect that the supported science community is responsive and accountable to the needs of the overall nuclear physics program and, via its scientific accomplishments, helps to shape the progress and evolution of the program.

Our recommendations have been guided by the following overarching principles and objectives:
- Ensure future excellence of the national program in nuclear theory and, on a broader scale, in nuclear science;
- Maximize the effectiveness of theoretical research and its contribution to the overall nuclear physics program;
- Develop the increased scientific workforce required to address the identified scientific opportunities and programmatic needs;
- Attract, educate, and retain the most talented scientists in the field;
- Reverse the decline of nuclear theory in top-ranked physics departments;
- Build program accountability into all major new program initiatives.

We are convinced that the following recommendations and initiatives, if followed and implemented by the agencies, will greatly improve the quality and effectiveness of the U.S. nuclear theory effort. Indeed, successful execution of the national scientific program in nuclear physics critically depends on these improvements.

## 5.2. Overview of initiatives and recommendations

We propose to achieve the principles and objectives described in Section 5.1 by a series of recommendations and initiatives, some of which require new funds or, in a constant level of effort scenario, redirection of funds from the existing base program. These new initiatives with such budgetary impacts are:

- Establishment of prize fellowships for postdoctoral researchers;
- Establishment of a fellowship program for graduate education;
- An increase in the funding level of Outstanding Junior Investigator awards;
- Establishment, by competitive bid, of Topical Centers with a specific programmatic thrust;
- Establishment, by competitive bid, of Centers of Excellence with generally interdisciplinary breadth;
- Investment in state-of-the-art computing facilities for lattice gauge theory, supernova simulations, and other CPU-intensive computational efforts.

In their efforts at implementing these initiatives, we urge the funding agencies to ensure the continued support of the existing strong research groups at a constant level of effort, which is crucial for maintaining our competitiveness and leadership in nuclear theory. Many of the most innovative and important advancements in theoretical physics cannot be planned, and the preservation of a vital base program in nuclear theory is absolutely critical to the long-term future of the field.

Our recommendations with less severe budgetary impact which we believe will, nonetheless, strengthen the nuclear theory programs supported by the two funding agencies are:

- Elimination of the disparity between NSF and DOE in funding per PI;
- Increased use of bridge funding to help create new faculty and staff positions;
- Leveraged support for sabbaticals;
- Enhanced presence of nuclear theory at top-ranked universities.

Several of our recommendations and initiatives require the annual competitive evaluation of applications by individuals or institutions. We recommend that, in addition to written peer review, the selection processes should include advice from independent committees of highly recognized scholars. Because of the substantial workload involved, we propose that one committee (with staggered two-year terms) would select the graduate fellows, a second (ad-hoc) committee would make recommendations about center proposals, and a standing committee with staggered three-year terms (see Section 5.3.1) would be established to select fellowship winners, provide advice on OJI awards, and evaluate proposals for bridge positions outside centers. We note that the members of such a standing committee could also help generate interest at leading academic institutions in creating new faculty positions in nuclear theory.

In the following subsections, we first discuss those new initiatives which require either additional resources or the reallocation of existing resources, followed by the recommendations with less severe budgetary impact. In Section 6 we make recommendations for the phase-in of the new initiatives in several budgetary scenarios, as requested by our charge. A detailed and comprehensive benefit analysis of our recommendations and proposed initiatives is presented in Section 7.

## 5.3. Fellowships and awards

### 5.3.1. *Postdoctoral prize fellowships*

We recommend the introduction of a national prize fellowship program (possibly called the "Weisskopf Postdoctoral Fellowship") for postdoctoral researchers in nuclear theory. Winning a prestigious fellowship in a national competition will raise the profile of a research career at an early stage and enhance the visibility of the brightest among our young scientists in the academic world. Giving the winners both support and freedom as they launch their research careers will maximize the scientific impact of these future leaders of the field at the crucial time when their abilities are fully developed and their energies are devoted solely to research. Furthermore, the existence and visibility of such a program will serve to attract highly talented students to do graduate work in nuclear theory. We recognize that the DOE will ultimately need to make the logistical decisions that define the Weisskopf Postdoctoral Fellowship precisely; our recommendations that follow are intended to convey how we think the initiative should be shaped to make it most effective.

The program would invite applications from young theorists not more than three years past their Ph.D., who would be free to choose their host university or national laboratory. We propose that up to five awards, with a duration of three years, should be made each year for a steady state of up to 15 fellowships. We recommend that the fellowship provide a stipend of about $50K plus benefits, and an annual allowance of $5K to be used at the fellowship holder's discretion for travel, computing, or hosting visitors. The fellowships should be awarded in December, before other postdoctoral positions are offered.

The fellowship procedures should be designed to achieve a balance between giving the winners freedom to choose how they want to develop their careers while at the same time ensuring that they find a host institution that welcomes them and where they are mentored well. Our recommendation closely follows the procedures developed for the Hubble Fellowship. Each applicant specifies their first, second, and third choice of destination. The head of the department at the institution of first choice must review and endorse the application before it is submitted, offer to provide office space, name a faculty or staff mentor, and waive indirect costs on the fellowship. This endorsement is not a recommendation, and an institution may endorse any number of applicants. After the selection panel has ranked the applicants, if two among the top five candidates list the same first choice institution, that institution can either accept the higher ranked of the two, or it can accept both winners, but then will not be allowed to host a Weisskopf Fellow selected the following year. If a winner does not get the first choice, this algorithm is repeated for the second choice. This process balances the freedom of choice of the young scientists who win the fellowship, the desire to ensure that they choose destinations where they will thrive, and the desire in the community that an institution which gets "lucky" in one year be excluded the next year.

The agencies could also consider initiating an annual symposium at which current holders of the fellowship give talks. Such a symposium, patterned on NASA's successful Hubble Fellows Symposia whose webcasts are often viewed by faculty search committees, would further augment the visibility of the fellowship and of nuclear theory, as well as building a sense of community among young leaders working in diverse areas of nuclear theory. The symposium could be permanently hosted by the INT or its location could rotate, for example, following either the DNP fall meeting or the national nuclear physics summer school.

A successful postdoctoral prize fellowship program - where success means that its recipients are seen by the broader physics community to be doing outstanding research and continuing onward to successful careers - will have positive impact on many fronts: it will help to attract good students to the field, retain good students as postdocs, raise the visibility of nuclear theory, and assist those seeking to make the case within their departments or laboratories for hiring of faculty or staff in nuclear theory. These multiple positive impacts warrant implementing this initiative even in an overall constant level of effort funding scenario. In this scenario, we propose to balance the additional expense of the new postdoctoral fellowships by a modest reduction in core program support, a painful cut that we do not take lightly and do not advocate in growth budget scenarios.

The selection of winners of this fellowship who subsequently go on to great success is crucial to maximizing the impact of this initiative. We therefore recommend establishing a panel of nationally renowned theorists, broadly representative of all areas of nuclear theory, which is charged with selecting the recipients of these prize fellowships. In order to provide continuity and memory to the selection process, membership in this panel should be by 3 year term. The panel could also be asked evaluate new bridge position proposals and to help with the selection of the OJI awards in nuclear theory.

*5.3.2. Graduate fellowships*

We recommend the introduction of a Nuclear Physics Theory Graduate Fellowship (NPTGF temporarily; possibly called the Bethe, Feshbach, Goeppert-Mayer, or Wigner Graduate Fellowship) funded by the DOE. The NPTGF would identify and support the best graduate students in the nation who intend to pursue theoretical nuclear physics research. The main objective of this initiative is to attract the highest caliber students to study nuclear theory. We recognize that the DOE will ultimately need to make the logistical decisions that define the NPTGF precisely; our recommendations that follow are intended to convey how we think the initiative should be shaped to make it most effective.

We recommend that the NPTGF be awarded to individuals in the early stages of their graduate study who are planning full-time uninterrupted study toward a Ph.D. degree. We suggest that students in their first or second year of graduate study be eligible to apply, and that exceptional senior undergraduates whose commitment to studying nuclear theory is already manifest also be eligible.

We recommend that up to ten fellowships be awarded annually. The NPTGF would offer recognition and three years of support for advanced study in nuclear theory. The second and third years of the award should be contingent upon certification by the fellowship institution that the Fellow is making satisfactory progress toward a Ph.D. in theoretical nuclear physics. We recommend that the fellowship provide an annual stipend of $25k. Each fellowship winner's group or institution will be expected to support the tuition and fees, and should be free to supplement the stipend if they wish to. The NPTGF would also offer each Fellow a one-time travel allowance designed to permit full-time study or research at a domestic or foreign site for at least 3 continuous months. Such travel should be endorsed by the research advisor.

We suggest that students must be nominated by their head of department or group leader, and that each university only be allowed to nominate one student per year. A student who is nominated would be asked to provide undergraduate and graduate transcripts, GRE scores, two or three reference letters including one from a future research advisor or program director, and a statement of research interest that outlines a proposed research direction. Applications should be reviewed by a panel of nationally renowned scientists broadly representative of the nuclear theory community, charged with picking the best candidates. While the Fellowship will be awarded to a U.S. university, the research supervisor can be a National Lab scientist. Care should be taken in the selection process to be receptive to exceptionally good students located in groups which do not habitually attract students of this caliber, as one benefit of this fellowship program should be to bring visibility and recognition to outstanding students in out of the way places.

As the goal of the NPTGF is to attract the very best students into the field, not just to increase the number of graduate students in nuclear theory, we correspondingly propose that the graduate fellowship program be funded out of existing funds used for graduate students.

### 5.3.3. Outstanding Junior Investigator awards

We commend the DOE for introducing the nuclear physics Outstanding Junior Investigator (OJI) program. These awards have helped to document the outstanding achievements of its young recipients and have increased their recognition at their home institutions and throughout the community. We believe that it is important to make the OJI awards research effective and even more attractive by raising the award level to at least $100k/year, so that recipients can support a postdoc and possibly a graduate student. We recommend opening up the competition for OJI awards to recently hired staff members in tenure-track positions at national laboratories. We recommend no change in the number of awards per year.

## 5.4. Research centers

### 5.4.1. Overview

We recommend, as an initiative, the creation of two types of centers. The first type, called topical centers by us, would attract the interest of national laboratories as well as university consortia which have a program oriented outlook. The broader or interdisciplinary centers of excellence, on the other hand, would be especially attractive to the universities, but we strongly believe that national laboratories should be eligible to compete for them as well. However, new centers at national labs will need to be justified in the context of the comments made in Section 2.3 on the role of the theory groups in national laboratories.

All centers should be reviewed after five years, with the possibility of one competitive renewal for another five-year period. Our recommendation, described below, is that the agencies build up to a steady state in which about ten topical centers and three to five centers of excellence are in operation, with two to three new centers being established every year. We expect that many center proposals will include the creation of new faculty and staff positions in nuclear theory. In instances where such senior personnel are hired explicitly through a bridging arrangement with the university or laboratory, by the end of the first 5 year term of a center the bridging arrangement should have been completed. However, even a very successful center that is renewed beyond 5 years will eventually finish its term. At this time, scientists hired at the tenure-track or equivalent level in the absence of specific bridging arrangements would, with strong positive reviews, be expected to transition onto support from the base program. We anticipate that the centers will act as magnets for graduate students and thus help to increase the quality of and the available support for nuclear theory graduate students.

We emphasize that these centers are not to be viewed as competition to the Institute for Nuclear Theory, which has been extremely successful. The INT has a well defined role as a national resource for all nuclear theorists; the centers proposed here have the purpose to focus research on various important topics or to strengthen nuclear theory in specific local environments. The centers should also have a role in the training of young scientists (graduate students and postdocs) working in a subfield of the center's scientific activity. For example, a center could organize one or two dedicated schools for young participants similar to the European Advanced Training Programme.

We consider the increased funding of centers as an important mechanism providing the necessary resources for an expanded faculty and staff bridge program. In this sense, as well as by virtue of their impact on the broader community, the proposed centers should function as "seeding centers" for the nuclear theory program.

*5.4.2. Topical centers*

Many of the critical needs of the program are currently not adequately covered within the DOE and NSF funded nuclear theory effort. A possible list of topics, which could be effectively addressed in topical centers, includes (the ordering is not intended to convey priority; the list is in no way meant to be complete):

- systematic development of EFT description of nuclear forces;
- properties of nuclei far from stability;
- microscopic study of nuclear input parameters for astrophysics;
- calculation of electroweak corrections to precision data;
- microscopic nuclear reaction theory;
- analysis of the spectrum of excited baryons (and mesons);
- phenomenology of hard probes of hot, dense matter;
- phenomenology of thermal probes of hot matter;
- simulations of core collapse supernovae;
- lattice simulations of hadron properties;
- lattice simulations of thermal QCD;
- ab initio many-body calculations;
- phenomenology of neutrino oscillations.

In order to successfully carry out the science of the Long Range Plan, an adequate theory effort is required for, not just a few of these topics, but - in principle - all of them. It is essential to involve theory groups at national laboratories in the pursuit of these tasks, because many of these problems require long-term, sustained research efforts that are often difficult to maintain in academic environments. Also, in many cases, individual university departments may not be willing to invest a sufficiently large number of faculty positions in these areas to create an effort with critical mass.

We propose that the agencies issue an annual request for proposals for such topical centers, which will be open for competitive bids by national laboratory and university groups. Proposals should be required to have a clear description of their relevance to the goals of the national nuclear science program, and they should contain a list of "deliverable" results anticipated during the award period. The centers would be expected to function as hubs of a wider network of scientists dispersed over various institutions (a focusing role), should have funds to support a sustained interaction and collaborations within this network, and should provide a vehicle to seed new faculty and staff. Care should be taken in the selection process to ensure that, particularly during the initial years of the program when the number of topical centers is steadily ramping up, the centers are distributed among different areas of nuclear theory in a way that takes into account scientific needs and opportunities.

At least 2-3 topical centers, with funding in the range of $300-500k each, should be established annually, up to a steady state of about ten such centers. National laboratories should be allowed to attract up to two of these centers, universities generally not more than one. This program would cost between $3M and $4M and inject new, but targeted funds up to $1M into participating theory groups at national laboratories or up to $500k into a strong university group. Each group would then have specific program responsibilities for which they can be held accountable by the funding agency through reviews.

We anticipate that, depending on its size, this program would generate between 30 and 50 new positions at the staff, faculty or postdoc level, leading to a considerable improvement of the pipeline for graduate students coming out of nuclear theory groups, as well as a rejuvenation of theory groups at the national laboratories and universities.

*5.4.3. Centers of excellence*

The second part of our centers initiative, broad and interdisciplinary centers of excellence, would target research areas benefiting from intense interaction with scholars from other communities, such as astrophysics, condensed matter physics, atomic physics, high energy physics, and computational science. These centers of excellence are intended both to rejuvenate nuclear theory and to promote its growth beyond traditional boundaries. We expect this type of center, with an intellectually broad and curiosity driven research agenda, to be especially appealing to university communities and national laboratories which can provide a fertile environment for interdisciplinary research projects. We envision these centers playing a multifaceted role, increasing the size and quality of the national nuclear theory effort, attracting outstanding talent into nuclear theory, promoting the development of nuclear theory in new directions, and highlighting our best work to the larger physics community.

A part of the grant support for such centers could be used to bridge new faculty and staff positions in nuclear theory. The ability of proposed centers to attract and retain high qual-ity graduate students and postdocs doing nuclear theory should be an important selection criterion. A general benefit of such centers of excellence is that they can provide significant leverage by motivating the institutions competing for them to contribute matching support in various ways. Another goal of the centers of excellence initiative is to elicit the interest of highly rated physics departments that currently have strengths in areas related to nuclear theory, which could be complemented by adding nuclear theorists to the faculty.

We believe that it will be important to allow for a wide range of possible structures without tight restrictions at the proposal stage, ranging from the NSF model of Physics Frontier Centers to centers of excellence that train students and postdocs in certain areas of need. The nature of the breadth that centers of excellence provide should also be left open, to allow the maximum creativity and local support and leverage from institutions involved in crafting specific proposals. Depending on their scope and thrust, such centers would require funding levels from $0.5 - $1M per year. With a total annual investment of $3M (in our most optimistic scenario), 3-5 such centers could be established.

The idea of soliciting proposals for such centers has been discussed in the community previously, but in our view the creation of such centers is now particularly timely and important, given the much wider scope of nuclear theory and its growing overlap with disciplines such as astrophysics, particle physics, atomic and condensed matter physics, as demonstrated in the achievements and opportunities sections of this document.

## 5.5. Computing

Some key nuclear theory problems can only be solved through numerical simulations on multi-Teraflop scale computational facilities. As outlined in the section on opportunities, understanding the confinement of quarks, the structure and spectrum of hadrons, and the phases and properties of hot QCD matter requires lattice QCD calculations; understanding the mechanism by which supernovae explode as well as the origin of the elements in these explosions requires multidimensional simulations; and solving the quantum many-body problem for nuclei requires large-scale quantum Monte-Carlo calculations. The 2002 NSAC Long Range Plan states that "theoretical work on these questions is crucial to obtain the full physics potential of the investments that have been made at Jefferson Lab and RHIC and new investments that are recommended for RIA and the underground neutrino physics lab. Advances in computational physics and computer technology represent great opportunities for breakthroughs in nuclear physics and nuclear astrophysics."

This is a uniquely opportune moment for our funding agencies to invest aggressively in computational nuclear science with the goal of solving problems of core importance to the physics program. The needs include both personnel and computer facilities. While personnel should be addressed through other initiatives, such as centers and fellowships, and through the base program, the critically needed computational infrastructure must be addressed separately. The required computational power for the next 3-4 years is on the order 10-15 Teraflops sustained. The investment in state-of-the-art computing facilities must be complemented by the programmatic support of strong user groups. Lattice QCD, supernova simulation and quantum many-body calculations all provide examples of possible topical centers. Whether via new topical centers or within the current base program, it is important that the agencies nurture strong collaborations of theorists who can take advantage of the capabilities offered by computing facilities to solve important problems central to the physics program, maximizing the realization of scientific opportunities.

Because of the scale of resources required and the fact that architectures optimized explicitly for lattice QCD are approximately 30 times more cost effective than general purpose machines in major computer centers, cost optimized dedicated facilities should be built. While they address specific hardware needs of the QCD community, these facilities should take advantage of synergies by responding to broader needs of the entire nuclear theory program. The bulk of these dedicated facilities should be placed at, for example, JLab and BNL, but it would be valuable to locate a small portion of the hardware investment in satellite centers located within a few university groups that commit to attracting, supporting and training outstanding graduate students and postdocs. We urge the funding agencies to actively seek new resources, such as SciDAC or new initiative money to fund these facilities. On the other hand, the opportunities are so compelling that even if new external resources cannot be secured and the funding is flat, a minimal sum of $3M/year from the nuclear science budget should be allocated to hardware investment. The nuclear theory program should contribute its appropriate share, as we discuss in Section 6. Based on current prices, this level of steady state funding would allow the acquisition of at least 10 Teraflops of computing power over 3-4 years.

The initial hardware investment in 10-15 Teraflops dedicated facilities would bring the US nuclear lattice QCD community to a competitive position with countries with advanced lattice QCD resources such as Germany, Japan, England, and Italy, and would make a number of important calculations, as outlined in Section 4.1, possible. Teraflops-scale machines are also necessary to the development of multidimensional supernova models with realistic neutrino transport and microphysics, and to reinforce our leading position in quantum many-body theory. As these internationally competitive resources are being established, it is essential to begin the long-term planning and commit to the long-term sustained investment needed to ensure that these facilities stay at the leading edge of the technology curve, thereby keeping the US computational nuclear community in a world-leading position in lattice QCD, supernova simulation and quantum many-body physics in the era of Petaflops-scale computers.

Currently, the US nuclear lattice community is working closely and actively with the high-energy lattice community to develop hardware and software resources. The synergy in using the same software and sharing the same QCD lattice configurations to accomplish their distinct physics objectives is a great benefit to both programs. However, it is also important to recognize the fundamental difference in physics interests of the two communities. The main focus of the particle physics lattice program is on extracting parameters of the standard model by calculating weak decay matrix elements, not on QCD per se. In contrast, the central goal of the nuclear physics effort in this area is to better understand how QCD works and how it gives rise to the fundamental features of strong interactions, of hadron structure and spectroscopy, of the phase diagram of QCD, and indeed of hadronic matter in all its manifestations.

Computational efforts in nuclear theory are now having an important impact in nuclear physics. The extensive investments made in experimental facilities, and those we have recommended in theoretical nuclear physics, must be balanced by a corresponding initiative in computational facilities. The computing initiative we have outlined is a crucial part of advancing the entire nuclear physics program. The combination of personnel and hardware investments will generate measurable outcomes that are essential for the fundamental progress of the field.

## 5.6. Recommendations without major budgetary impact

### 5.6.1. Disparity between NSF and DOE

A significant gap has developed in the resources given to University NSF and DOE nuclear theory researchers, with DOE funded theorists receiving nearly 30-50% more on average than their NSF counterparts. This puts NSF awardees at a significant disadvantage in achieving their research goals and has the additional consequence of discouraging the best young theorists from seeking NSF support. In order to remedy this problem, we strongly urge the NSF program directors to raise the level of support for their best research groups to be on a par with comparable groups in the DOE program. We recognize that this measure will, most likely, initially cause a reduction of the number of supported researchers in the NSF program, but it will serve to increase the quality of the program at the margin and will make the NSF program more attractive and competitive both for established researchers and for the young nuclear theorists, who are submitting their first grant application. We are convinced that this is the best way for the NSF to ensure the future vitality of its nuclear theory program and to retain its competitiveness with other science programs.

### 5.6.2. Bridging positions

It is essential to keep nuclear theory groups at universities and national laboratories vital and vigorous. This requires that retiring or otherwise departing nuclear theorists at these institutions be replaced, and where existing efforts are below optimal size, new staff or faculty positions be created. We urge the funding agencies to provide bridging support when such an investment facilitates the creation of new faculty and staff positions. We note that the DOE has, in selected cases, provided funds to bridge faculty or staff appointments. This support has been provided in the context of regular grants or contracts, and we expect this practice to continue. However, we anticipate that bridge support for faculty or staff positions will also be part of the budgets of many of the centers proposed in Section 5.4; in fact, this is an aspect of leverage to be considered favorably in the comparative review of center proposals.

As practice shows, there is no single organizational/funding scenario that would fit all possible situations: matching funds for the bridges usually come from a university, a laboratory, a center, or through a research grant. Under joint arrangements, nuclear theorists are appointed to tenure-track university positions at universities, or to staff positions in national labs, on a cost-shared basis with the funding agency. The agency agrees to reimburse the university or laboratory for one-half of the salary and benefits of a faculty or staff member.

While special care could be taken to ensure the balance within the theory program, we believe that it is not desirable to develop new targeted bridging programs, but that the competition for faculty or staff bridge funds should be open to the entire nuclear physics community. For this reason, we do not propose a bridging program with its own budget, but consider bridging arrangements as a part of other budget components, including the base program and the seeding centers. It is important that this already practiced option be publicized more broadly and aggressively than it is at present, to attract the attention of potentially interested institutions. It is equally important that the decision process for such bridging support be transparent and based on expert advice from the nuclear science community.

### 5.6.3. *Student recruitment*

The future of our field depends upon attracting the best young people to become nuclear theorists. There are many ways in which to achieve this goal, but we believe that one of the best is to expose students to some of the exciting aspects of nuclear research while they are still undergraduates. Thus we urge the nuclear theory community, with support by the agencies, to establish additional REU-like summer programs and undergraduate physics conferences. In order to be successful in attracting new talent into the field, these programs should be broad in scope, covering many areas of physics. Wherever possible we recommend that nuclear theorists should organize or co-organize such ventures. Perhaps more importantly, we strongly urge theorists to become mentors in such programs. In this way, the existing excitement in our field can be conveyed to the participating undergraduate students, and so help to attract some of them to pursue a research career in theoretical nuclear physics.

### 5.6.4. *Top-ranked universities*

The challenge of creating nuclear theory efforts in top-ranked physics departments that do not currently have them is both difficult and critically important. It is an essential goal because, if it can be accomplished, it will bring more of the best graduate students in the country into the field, resulting in more progress toward realizing the scientific opportunities that we have described and in defining new opportunities. We do not know a strategy that would work across the board: the potential nucleation sites (for example, experimental nuclear physics groups or theorists in other fields whose research connects to nuclear theory) and activation energy barriers are different in each university. We hope that some of the initiatives we are proposing - e.g. the prize postdoctoral fellowships and the centers of excellence - will occasionally serve to stimulate interest in nuclear theory and creation of nuclear theory faculty positions in top-ranked physics departments which presently do not have nuclear theorists on their faculty. Our central recommendation, however, is less specific: we urge the agencies to recognize the importance of strengthening nuclear theory within top-ranked physics departments, to provide support for bridging arrangements when opportunities arise, and to be flexible when extraordinary or unusual avenues toward strengthening nuclear theory or maintaining existing strength at these institutions open up.

### 5.6.5. *Sabbatical support*

Most nuclear theory efforts are small ones, with one or at the most two investigators. This has the consequence that such researchers can become somewhat isolated from many of the exciting new developments in the field. As a partial remedy for this situation, we recommend that the agencies make matching support available for researchers who want to spend sabbatical leaves at institutions where they can learn new techniques or otherwise reinvigorate their research capabilities.

# 6. Budget Scenarios

## 6.1. Overview

In response to the charge from the agencies, we propose three growth scenarios for nuclear theory funding: $5.8M (+23%), $9.7M (+39%), $13.7M (+55%), called Options I - III, in addition to the constant level of effort scenario (see Table 7). These scenarios are in line with various recent congressional initiatives for Office of Science funding. We stress, however, that the full proposed enhancement could be realized even within a more modest, e.g. 15%, increase ($60M) of the DOE nuclear physics budget, if the agency is willing to set 20% of the increase aside for nuclear theory. We believe that such a decision would be commensurate with the guidance provided by the NSAC Long Range Plan.

Table 7 represents "steady state" funding after a period of ramp-up. In the cases of Options II and III, the initiatives should be phased in over a period of 3-5 years. For example, the three different options could be implemented as steps of a long-term enhancement of nuclear theory funding culminating in the realization of Option III. A graphical representation of the recommended budget changes in the four scenarios is found in Figure 2.

An important exception is the recommended investment in computing equipment, which requires a steady-state annual investment of $3M in order to keep the computational infrastructure up to date. As we explained above (see Section 5.5), it seems reasonable for the nuclear theory program to contribute an appropriate share to this investment, as shown in the budget table.

**Table 7: Budget Scenarios (M$ annually, in FY2003 dollars)**

| Initiative | Const Effort[#] | Option I | Option II | Option III |
|---|---|---|---|---|
| Postdoc Fellows | 0.6 | 1.2 | 1.2 | 1.2 |
| Topical Centers | 0.5[*] | 1.0 | 2.5 | 4.0 |
| Centers of Excell. | | 1.5 | 2.5 | 3.5 |
| Computing | 0.2 | 1.0 | 2.0 | 3.0 |
| Enhanced OJIs | 0.0 | 0.5 | 0.5 | 0.5 |
| Total initiatives | 1.3 | 5.2 | 8.7 | 12.2 |
| Reprogr. DOE | -1.3 | 0.0 | 0.0 | 0.0 |
| | | | | |
| Chg. DOE bdgt. | 0.0 | 5.2 | 8.7 | 12.2 |
| Chg. NSF bdgt. | 0.0 | 0.6 | 1.0 | 1.5 |
| | | | | |
| Current NT | 25.0 | 25.0 | 25.0 | 25.0 |
| Future NT budget | 25.0 | 30.8 | 34.7 | 38.7 |
| | | | | |
| Increase (%) | 0 | 23 | 39 | 55 |

[*] In the constant level of effort scenario we propose to establish either a center of excellence or a topical center with a funding level of about $0.5M/year.

[#] Although we propose the introduction of a named graduate fellowship program, we do not recommend an overall increase in support for graduate students in the constant level of effort scenario. We anticipate a significant increase of total graduate student support in the growth scenarios as part of the center budgets.

## 6.2. Constant level of effort scenario

Several comments are in order concerning the recommended reprogramming of $1.3M of the DOE budget in the constant level of effort scenario. First of all, we need to point out that, apart from $0.9M dedicated to the INT workshop and visitor program, the entire nuclear theory budget goes to support the "base program." In a constant level of effort scenario any new initiative will require the reprogramming of funds now supporting the base program. Secondly, the three initiatives recommended for funding under a constant level of effort scenario – the postdoctoral and graduate fellowships, a single "seeding" center, and a very modest contribution to the computing initiative – broadly benefit the activities of the program. In fact, they take funds from the "base program" and reinvest them into it, but with an added level of visibility, leverage, and anticipated effectiveness.

We believe that it is important to initiate these new program components even under a constant level of effort scenario, for two main reasons: these initiatives are the first critical steps towards greater effectiveness and thereby support the experimental programs, and their implementation will provide additional focus and vitality to nuclear theory in the United States.

Because funded PIs and co-PIs have a certain flexibility in adjusting the spending of their grant support, we do not recommend specific line items for reductions in the constant level of effort scenario. Our expectation is that the cuts should be made in such a way that the strongest and most productive groups can at least retain their present level of effort. We are aware that the effect will be painful to many members of the research community, but the pain will be outweighed by the increased scientific effectiveness and vitality.

We would be remiss, however, if we were not to emphasize the negative impact that a constant level of effort scenario would have on nuclear theory and the nuclear physics program as a whole. Without an increase in the nuclear theory personnel, a large part of the theoretical research "opportunities" (more correctly: research needs) outlined in Section 4 will not be completed in a timely manner or not at all. This would be as detrimental to the national program as the failure to operate the existing world-class research facilities at full capacity, since it would seriously reduce the return on the investment in these premier scientific installations. In order to avoid such an outcome, the agencies – if faced with an overall constant level of effort scenario – should seriously consider allocating a larger fraction of their nuclear physics budget to the support of theoretical research.

## 6.3. Growth scenarios

As analyzed in the following section, the initiatives made possible in the growth scenarios would bring large benefits for theoretical nuclear physics research in this country. Option III would allow the agencies to support a personnel increase the by about 40%, make it possible to build an adequate theoretical base for the research required to reap the scientific opportunities offered by CEBAF and RHIC, and to develop the theoretical infrastructure for RIA. In addition, Option III would allow the nuclear theory budget to make the necessary investments in computational hardware from its own budget, without relying on taxes on the entire nuclear physics program.

Neither Option I nor Option II is optimal from the point of view of the national nuclear physics program. While some additional progress could be made toward addressing the many important scientific problems that need to be solved theoretically in order to interpret the data from the presently operating experimental facilities and to lay the groundwork for the next generation of facilities, the inadequate size of the theoretical workforce would continue to hold the national program back, as it does now. For this reason the agencies may consider it a prudent investment to realize Option III even when the growth of the total nuclear physics budget is less than 15%.

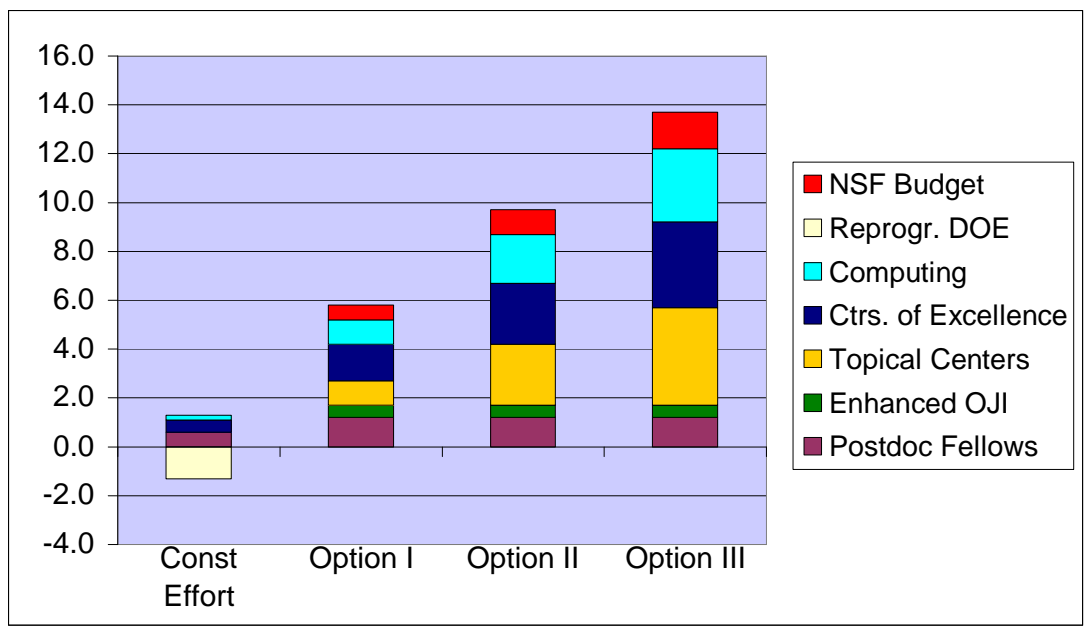

**Figure 2: Budget changes in the 4 scenarios (M$ annually)**

## 6.4. Priorities

As we explain in the following section, the benefits of our initiatives and other recommendations interact in manifold ways amplifying each other. For this reason, we believe that it would be inappropriate to arrange our recommendations into a linearly ordered, prioritized list. What is meaningful, from the standpoint of a benefits analysis, is to ask what the best uses for the overall program would be under a given budget constraint. This is what we have done, and what is encoded in our four budget scenarios. One can think of this as a "priority list," albeit a somewhat unconventional one: each priority level is composed of a bundle of items which, taken together, optimize the use of the available funds.

The items identified for implementation under a constant level of effort scenario are the "must do" items, corresponding to the highest ("zeroth") level of priority. If the nuclear theory budget at the DOE would grow by $5.2M, the bundle of initiatives called Option I would, by our judgment, provide the optimal investment from the view of the overall program, and so on. We do not believe that there would be similar benefit, if this additional amount of funds were allocated to a few initiatives only, such as topical centers and postdoctoral fellows, in order to implement them at the full recommended level shown under Option III. For the benefit of the reader, we present the bundled priorities as an ordered list:

1. *Constant level of effort* ("must do") *items:* half of the recommended postdoc and graduate fellowship programs, a single center, the full computing initiative (with only a token contribution from the nuclear theory budget), adjustment of the NSF grant size, bridging support as possible within the budget constraints;

2. *Option I:* the full recommended postdoc and graduate fellowship programs, two topical centers and centers of excellence each, enhancement of the OJI awards, the full computing initiative (1/3 contributed from the nuclear theory budget), adjustment of the NSF grant size, $0.6M increase in the nuclear theory budget of the NSF, increased bridging support as possible within the budget constraints;
3. *Option II:* the full recommended postdoc and graduate fellowship programs, six topical centers and three centers of excellence, enhancement of the OJI awards, the full computing initiative (2/3 contributed from the nuclear theory budget), adjustment of the NSF grant size, $1M increase in the nuclear theory budget of the NSF, increased bridging support;
4. *Option III:* the full recommended postdoc and graduate fellowship programs, ten topical centers and five centers of excellence, enhancement of the OJI awards, the full computing initiative (all paid by the nuclear theory budget), adjustment of the NSF grant size, $1.5M increase in the nuclear theory budget of the NSF, increased bridging support.

We again emphasize that our recommendation is to realize the scenario of Option III. Only this scenario would enable a level of theoretical activity as is required to meet the needs of the science program described in the 2002 NSAC Long Range Plan.

# 7. Benefit Analysis

One of the major considerations in the recommendations that are presented in this report is the possible impact that each recommendation would have on the field of nuclear theory, and on the community of nuclear theorists. The responses of the community to the questionnaire were taken into account, and the possible benefits and ill effects were carefully weighed. In the next few paragraphs, we outline what we see as the benefits of the recommendations made in this report. As a general comment, we stress the impact of both highly competitive fellowship programs and programmatic centers on the visibility of our subfield within the physics community and even the broader public. This enhanced visibility will be of benefit not only to the scientific community but also to the funding agencies themselves, which are always under pressure to document the effectiveness of their investments in research.

## 7.1. Fellowships and Awards

The clearest need identified in the questionnaire was that of additional personnel, in order to address the exciting and important scientific questions in an effective and timely manner. Many respondents stated that an increase in the number of postdoctoral researchers was the most important need. The proposed postdoctoral fellowships are aimed at addressing this specific need identified by the community, with the obvious benefit that effective and timely research could be carried out by these fellows and their mentors.

It is worth emphasizing that manpower is not just a matter of increasing numbers, but also of attracting and supporting students and postdocs of the highest caliber. New research opportunities attract highly talented and visible young researchers, the rising stars of the field, who are often in postdoctoral or pre-tenure positions. Enhancing this young cohort, from whom innovations often originate, will also improve the age demographics of the field. In addition, these fellows would constitute a highly visible talent pool for tenure-track positions at universities and staff positions at national labs. Furthermore, it is the opinion of the subcommittee that named and competitively awarded postdoctoral fellowships would serve as incentive for graduate students entering the field, in that they would highlight the scientific and career opportunities that can be attained after graduate education in nuclear theory.

The proposed graduate fellowships would serve a similar purpose. A number of well-supported, prestigious fellowships should provide a means of attracting some of the best talent into our field. Furthermore, such fellowships will ultimately disencumber some of the mentor's research funds, perhaps allowing the support of other graduate students, thus enlarging the pool of researchers. This would serve to increase not only the number, but also the quality of graduate students being trained in the field. It should also be noted that such fellowships can be used to increase the visibility of nuclear theory within a department, as well as the visibility of our field within the general physics community.

The proposed increase in the size of the OJI awards will also serve to increase the pool of researchers, if the awardees could support a postdoctoral researcher, for instance. This, together with the postdoctoral and graduate student fellowships proposed, will have the effect of increasing the number of researchers in the field at the pre-tenure-track/staff level, as well as the retention rate of the number of graduating students who go on to postdoctoral positions. Our aim is not simply to increase the numbers, but, more importantly, to enhance the quality of the graduate students and postdoctoral researchers. It must also be emphasized that increasing the retention rate after graduation is of crucial importance, as not only is personnel lost when students trained in the field do not stay in the field, but important expertise is also lost to nuclear theory. The recommendations made above should certainly help address the personnel needs identified by the community both in terms of quantity and quality and by helping to stem the "brain drain" of recent graduates leaving the field.

## 7.2. Centers

There are two main benefits to the centers of both types. The first is that more effective research will be carried out. It should be clear that what is envisioned here is not simply a number of researchers receiving further support to continue doing what they are already doing. What is expected is that there will be new synergy, a significant amount of "value added," so that in any of the centers the whole would be significantly greater than the sum of its parts. This would translate into more effective and focused research than the group of researchers could carry out without the center, and to significant progress made in the particular subject of their research. Furthermore, such centers should attract visitors interested in the research being carried out, thus stimulating the field more broadly than the researchers directly involved in the center could individually.

The second major benefit of the centers is that they are expected to serve as seeding centers for new faculty or staff positions in nuclear theory. Here, we mention one particular model, the Physics Frontier Center program run by the Physics Division at the National Science Foundation. In that program, there is significant cost-sharing required of the institutions hosting the centers, and this can take the form of new tenure-track positions. While we leave the details of the program to the agencies, we expect that a model like this will be followed, thus creating an increased scientific workforce. In addition, because of the intellectual activity associated with centers of either type, we believe that they will serve as excellent sites for the education of graduate students, and the continued education of postdoctoral researchers. In fact, we anticipate that a significant portion of the funding for centers will go toward support of graduate students and postdoctoral researchers, and this clearly helps to address the personnel needs of the field. It should also be noted that any center sited at a national laboratory, especially one with no large programmatic facility on site, will provide the much needed impetus and rationale for DOE to invest new resources in the intellectual infrastructure.

Less tangible but still important benefits will arise as institutions prepare proposals for submission to the competition for topical centers or centers of excellence, even before any centers are awarded. As proposers optimize their proposals they will discover synergies and interconnections within their institution or with other institutions. Furthermore, in making the case to their administrations for bridged positions in nuclear theory, they will be making the case for nuclear theory. These intangible benefits will bear immediate fruit in those instances where a center is awarded, and will bear fruit in the longer term in all places where proposals are prepared. For these reasons and because of the direct benefits of centers that we have described above, we recommend that even in a constant level of effort budget scenario, the DOE conduct a competition which results in the establishment of either one topical center or one center of excellence. If, as we expect, many outstanding proposals are received, this should help the DOE by providing evidence for the value of ramping up the number of centers and the nuclear theory budget in subsequent years to that we advocate in our optimal scenario.

## 7.3. Computing

Computational methods have become an integral part of, and an essential ingredient in, research in nuclear theory. A number of efforts outlined in the *Opportunities* section of this report depend crucially on the existence of adequate large-scale computational infrastructure. The obvious benefit to the field would be that of bringing these efforts to fruition in a timely manner. Conversely, if the required infrastructure is not provided these efforts will be severely hampered, perhaps even crippled, and all of the invested effort and time will be lost. Most importantly, many of the exciting questions that these collaborations seek to answer will remain unsolved, just as the efforts are promising tantalizing hints of what might be.

## 7.4. NSF Grants

This recommendation will undoubtedly cause apprehension throughout the pool of NSF-funded researchers, and will lead to some researchers being denied funding at the NSF, in the short term. However, in the long term, the effect should be increased funding for NSF Nuclear Theory. Programs at that agency grow only in response to extreme `proposal pressure': if there are exciting research opportunities being missed for lack of adequate funding in the program. This recommendation, if implemented, will create such `proposal pressure', which will hopefully stimulate significant growth in the funding of nuclear theory at the NSF. This adjustment will certainly make the NSF nuclear theory program more attractive to young researchers submitting their first grant applications. Without such new, young talent to invigorate it, the quality of research supported by this program will ultimately suffer, and it will fade into obscurity. One obvious benefit will be that the top researchers would be funded at a level that would allow them to function more effectively. Finally, if the funding for the program is increased in response to the proposal pressure, the obvious benefit would be that this program would be better able to contribute significantly in providing for the workforce and education needs of nuclear theory in the nation.

## 7.5. Summary

The initiatives recommended in this report are aimed at addressing the major needs identified by the community of nuclear theorists. The single most pressing need was that of additional research personnel, and we believe that this need will be addressed directly by the postdoctoral fellowships and the graduate student fellowships. Indirectly, the increased OJI support, and the creation of Topical Centers and Centers of Excellence will also aid in addressing this need, as will increasing the level of the NSF awards in nuclear theory. In addition, these initiatives will allow supported researchers to function more effectively than they now do. The Centers of Excellence will generate intellectual stimulation and excitement, while the Topical Centers will allow timely progress to be made on key questions. Taken together, we believe that the initiatives we propose will play a significant role in helping to realize the full scientific potential of the investment in the major experimental nuclear physics facilities.

# 8. Appendix I: Charges and Procedures

## 8.1. The NSAC charge

NSAC is asked to review and evaluate current NSF and DOE supported efforts in nuclear theory and identify strategic plans to ensure a strong U.S. nuclear theory program under various funding scenarios.

Among the opportunities and priorities identified in the 2002 NSAC Long Range Plan is an enhanced effort in nuclear theory and a large-scale computing initiative. Further guidance is requested at this time of how available resources might be targeted to ensure that the needed theoretical underpinnings are developed to realize the scientific opportunities identified by the community.

Your report should document your evaluation of the present national program in theoretical nuclear physics and its effectiveness in achieving results in the science areas highlighted in the recent 2002 Long Range Plan. It should identify the scientific needs and compelling opportunities for nuclear theory in the coming decade in the context of the present national nuclear theory effort, and what the priorities should be to meet these needs, including the development of a diverse highly trained technological workforce.

For both the DOE and NSF programs your report should provide advice on an optimum nuclear theory program under funding scenarios of i) a constant level effort at the FY2004 Nuclear Physics Congressional Requests and ii) with the increases recommended in the recent NSAC long range plan. For these funding scenarios the priorities, impacts and benefits of the various activities should be clearly articulated in the framework of a strategic plan. Your assessment should take into account the differences in the programs of the two agencies, as well as the unique roles of university investigators, the DOE national laboratories, and the Institute for Nuclear Theory. We request that an interim report be submitted by September 2003 and a written report responsive to this charge be provided by November 2003.

| | |
|---|---|
| James B. Hunt | Raymond L. Orbach |
| Acting Assistant Director | Director |
| Directorate for Mathematical and | Office of Science |
|     Physical Sciences | Department of Energy |
| National Science Foundation | |

## 8.2. NSAC letter to subcommittee

Dear Berndt,

As you know, Ray Orbach, Director of the Office of Science at DOE, and John Hunt, Acting Assistant Director for the Division of Mathematical and Physical Sciences at the NSF, have charged NSAC to review and evaluate the current NSF and DOE efforts in nuclear theory, and to identify strategic plans to ensure a strong nuclear theory program for the coming decade. The review is predicated on the strong recommendation in the 2002 NSAC Long Range Plan for an enhanced effort in nuclear theory and a large scale computing initiative. The charge asks for guidance, under two funding scenarios, for how the agencies can best target available resources to optimize the realization of the scientific opportunities identified by the community. The detailed wording of the charge, which I have previously forwarded to you, gives further instructions. The deadline for a preliminary report is this September and the final report is due in November, 2003.

I am writing to formally ask you to serve as the Chair of an NSAC subcommittee to consider this charge and to report back to NSAC. The work of this subcommittee is extremely important since the Long Range Plan gives such strong support to an increased role for nuclear theory. This is the opportunity afforded our community to give input as to how the funding resources can best be used to enhance and improve nuclear theory in this country.

There will be an NSAC Meeting in the Washington, D.C. area on May 30. We may ask you to provide a short status report on the activities, plans, and procedures of the subcommittee at that time. I will inform you further of this when the agenda is fully worked out.

I realize that this task imposes an extra burden on you, and I just want to express in advance my real appreciation to you that you have agreed to take on this responsibility. I will be available to help you in any way I can and will attend the subcommittee meetings in an ex officio capacity.

Best regards,

Rick Casten
Chairman
DOE/NSF Nuclear Science Advisory Committee                April 16, 2003

## 8.3. Subcommittee meetings

The Subcommittee convened for three regular meetings at the National Science Foundation in Arlington, VA (May 29, July 11, August 7-8, 2003), with participation of program officers from the NSF and DOE. In addition, the Subcommittee communicated through a large number of telephone conferences and electronic mail.

## 8.4. Expert witnesses

The subcommittee sought advice on special issues from the following experts:

    Larry Cardman (Jefferson Lab – the Jlab bridging program)
    Wick Haxton (Institute for Nuclear Theory – INT issues)
    Larry McLerran (BNL – university faculty bridging programs)
    Anthony Mezzacappa (ORNL – core collapse supernova simulations)
    John Negele (MIT – lattice gauge theory)
    Vijay Pandharipande (U. Illinois – numerical methods in many-body theory)

We received written statements about plans for RIA theory efforts from Don Geesaman (ANL) and Konrad Gelbke (NSCL). We thank Jacek Dobaczewski (Warsaw) for a thoughtful analysis of the role of the INT. We also informally consulted various theorists from neighboring subfields of theoretical physics, who have contributed to nuclear theory in the recent past by their research and by educating graduate students who have become nuclear theorists.

# 9. Appendix II: Community Surveys

## 9.1. Questionnaire to theorists

The following questionnaire was sent to all nuclear theorists currently supported by a grant or contract from the DOE or NSF. Principal investigators were asked to distribute copies to all co-PIs on their grant. We received 79 more or less complete responses to the questionnaire responses, which were distributed to all members of the subcommittee and collected in a Website with restricted access. The members of the subcommittee did not fill out the questionnaire, because we did not want to "double count" our own opinions. The 79 respondents constitute about 40% of the eligible number of senior personnel presently supported by the agencies senior personnel presently supported by the agencies (179 in FY02, see Table 1 in Section 2).

We have relied heavily on the responses to questions **A1** and **A2** in our selection of recent research achievements and important opportunities for future research and thus do not give a detailed analysis of the responses to these two questions. Summary analyses of the responses are listed immediately after each question, with the exception of **A1**, **A2**, and **D6**.

PREAMBLE

The DOE and NSF have charged NSAC to undertake a review of the nuclear theory program in the United States. A subcommittee (Joseph Carlson, Barry Holstein, Xiangdong Ji, Gail McLaughlin, Berndt Mueller (chair), Witold Nazarewicz, Krishna Rajagopal, Winston Roberts, and Xin-Nian Wang) has been charged to assess the quality and effectiveness of the currently supported program and to identify strategic plans to ensure a strong national nuclear theory program under various funding scenarios. Our charge does not call for an evaluation of the strength of one research group relative to others, but rather for an overall assessment of the strengths, accomplishments, opportunities, and needs of the field as a whole. The full text of the charge can be found on the NSAC Website at:

http://www.sc.doe.gov/henp/np/nsac/mtg030603/DOE_NSF_Charge_Letter.pdf

As part of our effort we have been asked to request broad input from the nuclear theory community in the United States. This questionnaire is a first attempt to solicit this input, and we ask for your support and cooperation. The questionnaire is distributed by the funding agencies to the leading PI's on all active grants at our request. We ask all PI's to distribute this questionnaire to their co-PI's or other faculty participating in their grants. (A few questions have been identified as "leading PI's only" in order to avoid duplication of answers to statistical questions.) On all other questions we hope to receive responses from all individual members of research group who share in a common grant.

We ask that you provide concise and candid answers to our questions. All members of the subcommittee have agreed to treat the responses with confidentiality, and we will make every attempt to avoid the distribution outside the committee of any response in association with an individual's identity. If you think that a certain question does not apply to you, please, feel free to ignore it. If you don't have an opinion with respect to some of the questions, do not feel obliged to provide an answer. And if you think that we have omitted an important question or neglected an important subject, please, send us your opinion or feedback, anyway. Of particular importance to us are the questions about the science (section A), strategic issues (section D) and question B16 about the need for a nuclear theory town meeting.

Please send your response to the following email address: mueller@phy.duke.edu

We hope to hear from you and your colleagues soon, preferably well in advance of our first subcommittee meeting on May 29. However, later responses will also be appreciated. We thank you, in advance, for your time and cooperation.

For the NSAC Theory Subcommittee,
Berndt Mueller

A. SCIENCE

*1. What do you think are the few most important contributions to progress in the nuclear physics program as a whole over the past 7-8 years due to theory advances? Please include some advances that are not in your own subfield.*

*2. What are the 2-3 key theoretical questions that must be answered for progress to be made in your subfield in answering the questions defined in the most recent long range plan:*
    *What is the structure of the nucleon?*
    *What is the structure of nucleonic matter?*
    *What are the properties of hot nuclear matter?*
    *What is the nuclear microphysics of the universe?*
    *What is to be the new Standard Model?*

*3. Are these key questions being adequately addressed by theorists in the United States today? If not, what can or needs to be done to focus attention and make progress on these questions? (E.g. is increased manpower needed? Is increased computational power needed? Or are the resouces available now, but have not been sufficiently focused on these questions? Do new theoretical techniques need to be developed?)*

62 responses:

| | | |
|---|---|---|
| Are the questions adequately addressed: | Yes (11) | No (24) |
| Increased computational power needed: | Yes (8) | No (12) |

Resources not focused on important questions:        Yes (4)
New theoretical techniques need to be developed:     Yes (13)

For the most part, respondents to this question thought that some of the key questions were being addressed, but perhaps not in an optimal fashion. A variety of suggestions on how to ameliorate this situation were made, with the overwhelming suggestion (expressed in the majority of responses) being the need for more personnel in every sub-field. The bottom line is: no progress can be made until the workforce issue is resolved.

Several respondents addressed the question of computational power. 12 people expressed an opinion that there is no big need for large-scale computational initiatives ("especially if that inhibits growth in manpower"; "creative thought, and not brute force, is the path to progress.") Most of those responding in favor of an increased computational power (8 people) represent lattice QCD. Here "computer infrastructure is absolutely crucial to mounting an effort in lattice QCD that has international impact."

Everyone agrees that there is need for new theoretical techniques, and 13 respondents expressed their support explicitly. Interestingly, several respondents (hadrons, RHIC) expressed their support for phenomenology ("there needs to be change in perception among the theorists themselves, concerning the value of phenomenological analysis" and "numerical modeling" versus "pure theory") while one respondent believes that phenomenology is playing a negative role ("We don't need more people running cascades or fitting multipoles.").

*4. Does the theory effort in your subfield need the formation of, and special support for, alliances of several theorists at multiple institutions in order to make adequate progress?*

62 responses:

Yes (46 responses; 74% )    No (12; 19%)        Maybe/No opinion (4; 7%)

Increase travel funding: 5
Increase visitor and sabbatical funding: 3
Theory Centers: 2

The majority of respondents (74%) believe that collaborations are important or very important. The mechanisms identified for fostering collaborations are: increased travel funding (5 responses), increased visitor and sabbatical funding (3 responses), and creation of topical centers (2 responses). The need for large, or focused, collaborations was predominantly expressed by lattice QCD and nuclear structure (RIA) representatives while most of the rest were not advocating this possibility (smaller collaborations and meetings in theory centers satify their needs). The bottom line here was that a `one size fits all' approach is not only undesirable, but could have adverse effects on the community, as well as on the science.

*5. With a modest increase in available funding for nuclear theory, which kind of investment by the funding agencies would, in your opinion, have the largest impact on the field?*

57 responses:

The respondents named the following investments (the sum of the numbers exceeds the total number of respondents, because some answers named several items):

Support for students and postdocs (30)
More positions for nuclear theory (23)
Increase base program (8)
Support for travel, workshops, and schools (5)
Theory centers (5)
More computers (4)
Sabbatical fellowships (3)
Public relation efforts (1)

A large majority of respondents believe that a modest increase in funding in nuclear theory should go toward increased manpower (93% of respondents): funding for students and postdocs is considered most important (53%); named, prestigious graduate and post-doctoral fellowships will help. Almost equally important are new positions for nuclear theory (40%), both at the national laboratories and at the universities. Here, bridged faculty positions, and research awards like the OJI and CAREER awards were seen as means of attracting and retaining young theorists deemed to be needed. Only 14% of responses consider the idea of increasing the base program. There is a need for an increased support for collaborative research, i.e., support for travel, workshops, schools (9%), topical centers (9%), and sabbatical fellowships (5%).

Four respondents (7%), representing lattice QCD, thought that buying more computer power will have the largest impact.

As a more long-term goal, one respondent suggested that the community needs to reach out to the budding physicists in high school, by advertising the intellectual quality of the problems confronting nuclear theory. Increased and improved public relation efforts would serve to attract some of the best young students into this field.

B. PERSONAL AND GROUP DATA

*1. What is your "academic age" (years after the Ph.D. degree)*
   *<10    10-20    20-30    30-40    >40*

*2. Which field of physics best describes the activities of your thesis advisor?*

*3. Did you receive your doctoral degree from an institution outside the United States? If so, in which country?*

*4. Is or was your position partially supported by "bridge" funding from a laboratory or center, or through a grant?*

*5. For how many years have you been consecutively funded?*

*6. Give the percent effort of your research according to the Long Range Plan classification*
    *a. Hadron structure*
    *b. Nuclear structure*
    *c. Hot and dense nuclear matter*
    *d. Astrophysics*
    *e. Fundamental symmetries*
    *f. Interdisciplinary*

73 responses (B1 – B6):

We present a statistical analysis of the answers to these demographic questions in the table below. "Age" indicates years past Ph.D., # means number of respondents. The columns a-f give the percentage of activity assigned to the five areas of the Long Range Plan (a-e) and interdisciplinary research (f).

The number of respondents giving as field of expertise of their thesis advisor is: nuclear physics (35), high energy physics (21), many-body theory (5), hadron or intermediate energy physics (4), few-body theory (2), lattice QCD (2), nuclear chemistry (1), statistical physics (1), mathematical physics (1), atomic physics (1).

| "Age" | # | US | Abroad | Bridge | a | b | c | d | e | f |
|---|---|---|---|---|---|---|---|---|---|---|
| 00-10 | 14 | 6 | 8 | 8 | 9.7 | 21.8 | 43.2 | 1.4 | 8.9 | 15.0 |
| 10-20 | 21 | 15 | 6 | 6 | 14.0 | 25.0 | 32.1 | 8.3 | 11.4 | 7.6 |
| 20-30 | 15 | 9 | 6 | 5 | 28.9 | 31.7 | 16.5 | 10.7 | 2.3 | 8.5 |
| 30+ | 23 | 16 | 7 | 2 | 22.4 | 35.2 | 17.0 | 3.7 | 12.6 | 9.1 |
| Total | 73 | 46 | 27 | 21 | 18.9 | 29.0 | 26.3 | 6.0 | 9.4 | 9.7 |

*7. How many graduate students do you currently advise?*

*8. How many graduate students under your supervision have graduated during the past 10 years, and what is their employment situation now?*

72 responses (B7 – B8):

43 respondents wrote that they are currently supervising graduate students, for a total of 86.5 students. For comparison, the total number of graduate students supported by agency funds (FY02) is 128.

47 respondents replied that graduate students had received their Ph.D. degree under their supervision during the past 10 years, for a total of 159 students. The most productive respondents listed 11, 10, and 9 graduated students, respectively. For comparison, the DOE manpower tables state that 263 Ph.D. degrees were awarded to supported students over the past decade.

*9. (Single and leading PI's only) How many postdocs are currently doing research in your group, and what is their fraction of grant support?*

38 responses:

30 respondents (we had asked that only PI's answer this question) wrote that their research group currently hosts at least one postdoctoral fellow, for a combined total of 73 postdocs. On average, research grants provide 65% percent of the support for these people. For comparison, the total number of postdocs supported by agency funds (FY02) is 80.

*10. What fraction of your effort is coupled to major experimental initiatives? Please quantify.*

68 responses:

The respondents stated that their research was closely related, in part, to a major experimental initiative, with an average fraction of 64% dedicated to such activities.

*11. Do you participate in outreach activities, such as K-12 education, recruitment of underrepresented minorities, summer research programs, mentoring, etc.? Please specify.*

72 responses:

Exactly half of the respondents stated that they participate in outreach activities.

*12. Have you participated in the INT or ITP programs or workshops during the past five years? If so, how?*

73 responses:

59 respondents reported that they had taken part in activities of the INT or ITP during the past five years, 14 stated that they had not.

*13. Have you participated in activities at the ECT\* during the past 5 years?*

72 responses:

50 respondents reported that they had participated in activities of the ECT\* during the past five years, 22 stated that they had not.

*14. Are you actively involved in international collaborations?*

72 responses:

56 respondents stated that they are actively involved in international collaborations, 16 wrote they were not.

*15. Are you involved in large-scale computing (indicate if you are a member of a larger collaboration)? If so, which infrastructure (e.g., NERSC, ORNL-CCS, Riken QCDSP, JLab SciDAC cluster, in-house workstation farm,...) provides you with computing resources? Are these resources adequate for your needs? What fraction of your research is dedicated to computational work?*

73 responses:

29 respondents stated that they are involved in large-scale computational efforts, 44 reported they were not.

*16. Do you see a need for a nuclear theory town meeting (or town meetings related to subfields) to provide input to the NSAC theory subcommittee beyond that we are currently gathering from the community via this questionnaire?*

71 responses:

47 respondents stated that they did not see the necessity for a town meeting, 17 wrote that a town meeting would be useful, and 7 expressed no opinion. Many of those who were not in favor of a town meeting expressed their view that the questionnaire was a much better means of soliciting input from the community at large.

*17. What is the most pressing issue facing your group at this moment? How can this issue best be addressed within a 15% increase in the nuclear theory budget?*

67 responses:

35 respondents gave a dearth of manpower as the most pressing issue for the field. 20 of these specifically stated a need for more postdocs; 7 favored new faculty or staff positions. 13 respondents wrote that the most pressing need was an increase in base fund-ing; 5 emphasized the need for bridge positions; 6 gave priority to graduate students; and 4 stated computing facilities as the greatest need.

C. INSTITUTIONAL ISSUES

*1. (Single and leading PI's only) How many faculty positions in nuclear theory have been filled at your institution over the past decade? How many of these were replacements and how many were new faculty lines? Has bridge funding been an important element in establishing these positions?*

39 responses:

A total of 23 institutions reported hiring a nuclear theorist within the past decade, 5 institutions reported no hires during that time frame. Of those that hired theorists, the average number of theorists hired was two. Many institutions hired one theorist, while several hired 4-5 theorists. Of the people hired 40% were identified as new positions, 40% replacements and 20% unknown. Approximately one-half of the new positions were supported by bridging and/or joint appointments.

*2. (Single and leading PI's only) Do you anticipate a search in nuclear theory at your institution over the next 5 years? Would the availability of bridge funding have a major impact on the chances of such a search to occur?*

39 responses:

Slightly over one-half of the institutions reported plans to conduct a search within the next five years. Of those responding to the question about bridge funding, many indicated that it would be valuable. Only two institutions reported that bridge funding would not be of interest.

*3. (Single and leading PI's only) If you do not anticipate a search in nuclear theory at your institution in the near future, give a reason or reasons why:*
   - *All slots are filled and we do not expect a retirement*
   - *My department/dean is not interested in nuclear theory*
   - *Other reasons.*

18 responses:

From this relatively small pool of respondents, slightly more than one-third said their department, head and/or dean had no interest. Another nearly equal group said that all slots were filled, while a significant number said that they had recently hired and that other groups were in greater need.

*4. Is there an experimental nuclear physics group at your institution? If so, are there active interactions between the theorists and the experimentalists?*

59 responses:

Roughly 90% percent of the respondents reported the presence of an experimental group at their institution, and 90% of these said there was active collaboration between experimentalists and theorists.

*5. Do you feel satisfied with the talent pool of graduate students at your institution who are interested in nuclear theory? If not, what are the biggest obstacles to attracting graduate students to your group?*

49 responses:

Only about 40% of the respondents are satisfied with their talent pool. Among those who are not, about one quarter cite lack of jobs, and another quarter lack of resources. The rest of the responses are split among the following: it is hard to attract students away from "hot" areas like string theory, it is hard to get students at a laboratory, and about 20-30% cite some aspect of the graduate program, e.g. the students in general are not of high enough quality or there are not enough nuclear physics courses.

D. STRATEGIC ISSUES

*1. What are the major strategic areas of nuclear theory which are vital for the overall program?*

55 responses:

As might be expected, there was no general consensus here, with much support being generated by and large for the field of the respondent. However, areas which received a good deal of general support included: RIA physics, nuclear astrophysics, RHIC physics, symmetries, and QCD. In detail, specific support was distributed among subfields: RIA/nuclear structure (14), RHIC/dense matter (9), Astrophysics (11), Effective Field Theory (5), Symmetries (8), Lattice QCD (5), Many-body physics (4), Neutrinos (4), QCD (6).

*2. What are the best vehicles for invigorating the field:*
   *a. Fellowship and scholarship programs?*
   *b. Bridged positions?*
   *c. Better grant support for junior faculty members?*
   *d. Other?*

70 responses:

There was a general consensus here that invigoration of the field could best be accomplished through bridge positions, as well as through some combination of fellowship and scholarship programs to support young students and postdocs. However, there was also a strong feeling that the best mechanism by which to achieve this aim is the existence of increased and reliable grant support for nuclear theory investigators - both junior and senior. Of course, it was also noted that the existence of jobs in the field is also a necessary ingredient.

Among the responses, support for (a) was expressed by 38, (b) by 41, and (c) by 40. The importance of jobs under (d) was mentioned by 14 respondents.

*3. How can we best keep nuclear theory vital at universities that currently have an active program?*

64 responses:

Again it was emphasized that the best way to retain vitality in the field was via the existence of positions, both bridge and other, and and through solid and reliable grant support. However, it was also generally felt that it was important to sell the field through inspiring talks and good teaching. Finally, it was noted that one of the best mechanisms to retain continued strength was through outstanding research!

Increased funding was emphasized by 31 respondents, more positions by 14, outstanding research by 10, and outreach and public relations by 11.

*4. How can we best stimulate nuclear theory research at universities that do not currently have a nuclear theory group?*

53 responses:

Many respondents noted that this is a difficult problem, saying little more.

21 respondents suggested bridge positions. Some among these warned to avoid creating places with one "lonely PI." 5 people explicitly disagreed with the idea of bridge positions, warning that it is not good to start new subcritical efforts, and advocated concentrating support on universities with current efforts.

10 respondents stressed the importance of interdisciplinary research, as part of selling the program, and of people doing work at intersections between nuclear theory and other fields who could help begin nuclear theory efforts at their universities. 3 people suggested that nuclear experimentalists could help at their universities.

5 people said that scholarship or fellowship programs would be of value, both by attracting good students or postdocs to join or stay in the field, and perhaps by getting nuclear theory postdocs into places with people doing work at intersection areas.

7 respondents wrote about the importance of publicizing discoveries, and giving colloquia. 6 people said that the best way is for people in the field to do nuclear theory that captures the imagination of those in other disciplines, and consequently also deans and provosts.

*5. What do you see as the most important contributions nuclear theory groups in the national labs can make to the national program? What could be done to further enhance the effectiveness of these groups?*

57 responses:

20 respondents stressed the importance of providing support, guidance, help with understanding theory behind experiments, and long term planning to the local experimental efforts. Several also noted that they should articulate the lab program to the outside community, and should lead the theory effort of the larger community. One person wrote that they "should not be inflexibly tied to the local experimental program."

17 respondents expressed the view that the laboratory groups should undertake long term, large scale, computationally intensive projects.

Many people stressed contact with and support for nuclear theorists at universities. In this vein, 9 advocated the importance of collaboration with theorists at universities, 8 stressed organizing conferences, workshops and schools and 12 stressed visitor and sabbatical programs, and 4 suggested that theory students should spend time at labs, and labs should be more involved in training students.

*6. What do you see as the role of nuclear theory in answering big intellectual questions (e.g., Turner report, URL = http://books.nap.edu/books/0309074061/html) and in serving the nation's needs. Please be as specific as possible.*

*7. Is the present model of funding satisfactory? If not, how should it be changed? Which opportunities would be missed if the present model of funding continues?*

60 responses:

37 respondents answered "yes," or "yes, but the level of funding must be increased," that is, the model of funding is OK but the level is not. 3 expressed "no opinion."

There was no consensus as to suggested changes to the model. We list here only those ideas mentioned by several respondents. Five people advocated broadening the funding agency definitions of nuclear theory, or breaking down the funding agency divisions between atomic/nuclear/CM/astro/particle theory, typically because of concerns that interdisciplinary research is neglected by the agencies. Four people advocated funding bridge positions and larger grants for young people. Three other respondents said that the present model is based too much on history: Unproductive people get flat-funded rather than cut and the field must be willing to cut funding to low performers.

*8. Should support for nuclear theory be more closely linked to major experimental initiatives?*

67 responses:

Yes (9): Many of these noted that intellectual independence, freedom and creativity should not be curtailed. An additional 14 answered "occasionally," "perhaps," or "in some instances." Examples mentioned included the need to support partial wave analysis at JLab, and the need to support theory for new projects like RIA. Several respondents answered "yes at national labs."

No (44): Many of these respondents noted the importance of linkage to experiment. Many stressed the importance of theory-experiment interaction, especially in the case of major initiatives. These respondents felt that the current degree of linkage is appropriate, and answered "no" because of the word "more" in the question. Many noted that one important role of theory is to guide experiments into new directions, something that would not happen if theory is too closely linked to existing experiment. Theory must always be connected to and inspired by experiments, but explorations in directions other than those of current experiments should be encouraged if they lead to a better understanding of nature.

*9. Are you satisfied with the role of the INT in stimulating and advancing research in theoretical nuclear physics? If not, which changes would you like to see?*

69 responses:

The majority of the respondents (over 70%) expressed their appreciation for the work of the INT and praised its role in stimulating and advancing research in nuclear physics. In the words of one respondent: "the INT is one of the best (if not the best) thing that has happened in nuclear theory in a long time." About 15% were not satisfied, with some complaining that the INT was not cost effective, and others are not sure. The most often made suggestion for adjustment of the program structure was that the INT could be more effective if it organized smaller and shorter programs and more workshops. Other respondents desired more programs in certain areas of physics, such as RIA related nuclear structure physics and hadronic physics, and had a closer linkage to major experimental programs.

*10. Should there be more emphasis on interdisciplinary aspects of nuclear theory research at the expense of focused research?*

65 responses:

Nearly half of the respondents said no, about 1/3 said yes, and the others either have no strong opinion or did not understand what the question meant. Those who said no argued that it would dilute the core program, that there was already enough interdisciplinary research in nuclear physics, and that it was too much a buzz word. Those who said yes argued that interdisciplinary research is one of strongest aspects of nuclear physics, for example, nuclear astrophysics, quantum many-body physics, and nuclear-particle physics, and that it opens up new directions and enhances nuclear physics. Some respondents suggested that individual researchers, and not the funding agencies, should decide whether they work on interdisciplinary problems.

## 9.2. Questionnaire to experimentalists

The following abbreviated questionnaire was sent to all PIs on experimental research grants and contracts from DOE or NSF.

PREAMBLE

The DOE and NSF have charged NSAC to undertake a review of the nuclear theory program in the United States. An NSAC subcommittee chaired by Berndt Mueller has been charged to assess the quality and effectiveness of the currently supported program and to identify strategic plans to ensure a strong national nuclear theory program under various funding scenarios. The charge does not call for an evaluation of the strength of one research group relative to others, but rather for an overall assessment of the strengths, accomplishments, opportunities, and needs of the field as a whole. The full text of the charge can be found on the NSAC Website at:
http://www.sc.doe.gov/henp/np/nsac/mtg030603/DOE_NSF_Charge_Letter.pdf

As part of their effort, the subcommittee has been asked to request broad input from the nuclear physics community in the United States. The leading PI's on all active DOE and NSF theory grants have already received a detailed questionnaire through their funding agencies. The remaining US-based theorists have been contacted via the DNP. This mailing solicits the input from the leading PI's on all active DOE and NSF experimental grants.

The questions contained in the brief questionnaire attached below are of general nature and aim at addressing a number of points that are vital to the field. All members of the subcommittee have agreed to treat the responses with confidentiality, and they will make every attempt to avoid the distribution outside the committee of any response in association with an individual's identity.

Please send your response to the following email address: mueller@phy.duke.edu

We hope to hear from you soon. We thank you, in advance, for your time and cooperation.

For the NSAC Theory Subcommittee,
Berndt Mueller

QUESTIONNAIRE

1. Please list your major subfield in accordance to the classification of the latest NSAC Long Range Plan:

    - Structure of the nucleon
    - Structure of nucleonic matter
    - Properties of hot nuclear matter
    - Nuclear microphysics of the universe
    - Beyond the Standard Model

2. What are the 2-3 key theoretical questions that must be answered for progress to be made in your subfield in answering the questions defined in the most recent long range plan? Are these key questions being adequately addressed by theorists in the United States today?

3. How can we best keep nuclear theory vital at institutions that currently *have* an active program? Do you anticipate a search in nuclear theory at your institution in the near future? If not, please, give a reason or reasons why:

    - All slots are filled and we do not expect a retirement
    - My department/dean is not interested in nuclear theory
    - Other reasons.

4. How can we best stimulate nuclear theory research at institutions that *do not currently have* a nuclear theory group?

5. Should support for nuclear theory be more closely linked to major experimental initiatives? If your answer is yes, please, elaborate which mechanisms might facilitate such closer links.

6. Please, share with us any other comments you have, which are relevant to our sub-committee's charge:

**Evaluation of Responses:**

We received a total of 22 responses from experimentalists to our questionnaire. These represent a very small fraction of the experimental community and one should keep in mind the possible inaccuracy in their reflection of the community. Among those who responded, 5 ¾ describe their major subfield as structure of the nucleon, 3 ¾ as structure of nucleonic matter, 4 as properties of hot nuclear matter, 3 ¼ as nuclear microphysics of the universe and 5 ¼ as beyond the Standard Model. 8 respondents listed more than one subfield.

The responses raised many theoretical questions that are considered critical for progress to be made in each subfield. Most of these questions are considered not adequately addressed or the development has been very limited. Some of the frequently mentioned questions are:

1. Structure of the nucleon: Understand nucleon structure in terms of quark and gluon degrees of freedom in QCD and connect to the experimental measurements to form a coherent picture of the nucleon.
2. Structure of nucleonic matter: Understand microscopic structure of complex and exotic nuclei in terms of fundamental strong interactions.
3. Properties of hot nuclear matter: Microscopic description of the dynamics of heavy-ion collisions and extraction of properties of dense matter from experimental data.
4. Nuclear microphysics of the universe: Understand the nuclear structure far from stability.
5. Beyond standard model: How the observed neutrino mass is generated and what experimental measurements are needed to elucidate?

A few responses are optimistic about having an opening in nuclear theory in their departments, though the current financial difficulties in public and private funding may further delay the process. Increased funding for the existing theory program will always help to bolster nuclear theory research and help attract good students. Bridged positions are always helpful to bring in young people. However, it will have a much larger impact if such position can be created in upper-tier departments. (The answer to whether you expect a new or replacement hire in your department is: 5 yes, 10 no, 2 maybe/difficult, 5 no answer).

The majority of respondents think that some theory support should be linked to experimental initiatives, but not completely tied to specific experimental problems. The interaction between theory and experiment are mutually beneficial. Many suggest formation of some close collaboration between theorists and experimentalists on problems that need theoretical guidance and phenomenological work.

# Glossary

| | |
|---|---|
| ANL | Argonne National Laboratory |
| BNL | Brookhaven National Laboratory |
| CPI | Consumer Price Index |
| DOE | Department of Energy |
| EFT | Effective Field Theory |
| ECT* | European Center for Theoretical Nuclear Physics |
| FY | Fiscal Year (October 1 – September 30) |
| GeV | Gigaelectronvolt (1 billion electron volts) |
| GRE | Graduate Record of Education |
| INT | DOE Institute for Nuclear Theory (Seattle) |
| ITP | NSF Institute for Theoretical Physics (Santa Barbara) |
| JLab | Thomas Jefferson National Accelerator Facility (TJNAF) |
| keV | kiloelectronvolt (1000 electron volts) |
| KR | Koonin Report |
| LANL | Los Alamos National Laboratory |
| LBNL | Lawrence Berkeley National Laboratory |
| LIGO | Laser Interferometer Gravitational Wave Observatory |
| MeV | Megaelectronvolt (1 million electron volts) |
| NPTGF | Nuclear Physics Theory Graduate Fellowship |
| NSAC | Nuclear Science Advisory Committee |
| NSCL | National Superconducting Cyclotron Laboratory |
| NSF | National Science Foundation |
| NUSL | National Underground Science Laboratory |
| OJI | Outstanding Junior Investigator |
| ORNL | Oak Ridge National Laboratory |
| QCD | Quantum Chromodynamics |
| Petaflops | 1 quadrillion floating point operations per second |
| PI | Principal Investigator |
| R&D | Research and Development |
| REU | Research Experience for Undergraduates |
| RHIC | Relativistic Heavy Ion Collider |
| RIA | Rare Isotope Accelerator |
| SciDAC | Scientific Discovery through Advanced Computing |
| SM | Standard Model (of Particle Physics) |
| SNO | Sudbury Neutrino Observatory |
| SCET | Soft Collinear Effective Theory |
| Teraflops | 1 trillion floating point operations per second |